\documentclass[lineno]{jfm}

\usepackage{graphicx}
\usepackage{newtxtext}
\usepackage{newtxmath}
\usepackage{float}
\usepackage{amssymb,mathtools}
\usepackage[arrowdel]{physics}
\usepackage[round]{natbib}
\usepackage{csquotes}
\usepackage{float}
\usepackage{tikz}
\usepackage{hyperref}
\hypersetup{
    colorlinks = true,
    urlcolor   = blue,
    citecolor  = black,
}

\newcommand{\RomanNumeralCaps}[1]
\nolinenumbers

\shorttitle{Resolvent analysis of turbulent jets in flight}
\shortauthor{I. A. Maia et al.}

\definecolor{color1}{RGB}{0,113,188}
\definecolor{color2}{RGB}{216,82,24}
\definecolor{color3}{RGB}{236,176,31}

\newsavebox\mycircle
\savebox\mycircle{                                                                   
  \begin{tikzpicture}                                                           
    \tikz\draw[line width=0.3mm, color=color1] (0.05,0.05)
circle (0.5ex); 
  \end{tikzpicture}                                                             
} 

\newsavebox\mycircleblack
\savebox\mycircleblack{                                                                   
  \begin{tikzpicture}                                                           
    \tikz\draw[line width=0.3mm, color=black] (0.05,0.05)
circle (0.5ex); 
  \end{tikzpicture}                                                             
} 

\newsavebox\myline
\savebox\myline{                                                                   
  \begin{tikzpicture}                                                           
    \tikz\draw[line width=0.3mm, color=color2] (0,-0.05)--(0.5,-0.05);
  \end{tikzpicture}                                                             
} 

\newsavebox\mylineblue
\savebox\mylineblue{                                                                   
  \begin{tikzpicture}                                                           
    \tikz\draw[line width=0.3mm, color=color1] (0,-0.05)--(0.5,-0.05);
  \end{tikzpicture}                                                             
} 

\newsavebox\mylineblack
\savebox\mylineblack{                                                                   
  \begin{tikzpicture}                                                           
    \tikz\draw[line width=0.3mm, color=black] (0,-0.05)--(0.5,-0.05);
  \end{tikzpicture}                                                             
}

\newsavebox\mylinedashed
\savebox\mylinedashed{                                                                   
  \begin{tikzpicture}                                                           
    \tikz\draw[dashed ,step=0.2, color=black] (0,-0.05)--(0.5,-0.05);
  \end{tikzpicture}                                                             
}

\newsavebox\mylinedashedblack
\savebox\mylinedashedblack{                                                                   
  \begin{tikzpicture}                                                           
    \tikz\draw[dotted ,step=0.2, color=black] (0,-0.05)--(0.5,-0.05);
  \end{tikzpicture}                                                             
}

\newsavebox\mylinedashedblue
\savebox\mylinedashedblue{                                                                   
  \begin{tikzpicture}                                                           
    \tikz\draw[dashed ,step=0.2, color=color1] (0,-0.05)--(0.5,-0.05);
  \end{tikzpicture}                                                             
}

\newsavebox\mylinedashedred
\savebox\mylinedashedred{                                                                   
  \begin{tikzpicture}                                                           
    \tikz\draw[dashed ,step=0.2, color=color2] (0,-0.05)--(0.5,-0.05);
  \end{tikzpicture}                                                             
}

\newsavebox\myboxblue                                                           
\savebox\myboxblue{                                                                   
  \begin{tikzpicture}                                                           
    \draw [line width=0.3mm, color=color1] (0.1,0.1) rectangle(0.3,0.3);                                       
  \end{tikzpicture}                                                             
}

\newsavebox\myboxred
\savebox\myboxred{                                                                   
  \begin{tikzpicture}                                                           
    \draw [line width=0.3mm, color=color2] (0.1,0.1) rectangle(0.3,0.3);                                       
  \end{tikzpicture}                                                             
}

\newsavebox\myboxblack                                                           
\savebox\myboxblack{                                                                   
  \begin{tikzpicture}                                                           
    \draw [fill=black] (0.1,0.1) rectangle(0.3,0.3);                                       
  \end{tikzpicture}                                                             
}  

\newsavebox\myboxgray                                                           
\savebox\myboxgray{                                                                   
  \begin{tikzpicture}                                                           
    \draw [fill=gray] (0.1,0.1) rectangle(0.3,0.3);                                       
  \end{tikzpicture}                                                             
}   

\newsavebox\myboxwhite
\savebox\myboxwhite{                                                                   
  \begin{tikzpicture}                                                           
    \draw [fill=white] (0.1,0.1) rectangle(0.3,0.3);                                       
  \end{tikzpicture}                                                             
} 

\title{The effect of flight on a turbulent jet: coherent structure eduction and resolvent analysis}

\author{Igor A. Maia\aff{1}
  \corresp{\email{igoriam@ita.br}},
  Liam Heidt\aff{2},
  Ethan Pickering\aff{3},
  Tim Colonius\aff{2}
  Peter Jordan\aff{4}
  \and Guillaume Br\`es\aff{5}}

\affiliation{\aff{1}Divis\~ao de Engenharia Aeron\'autica, Instituto Tecnol\'ogico de Aeron\'autica, S\~ao Jos\'e dos Campos,12228-900, Brazil
\aff{2}Division of Mechanical and Civil Engineering, California Institute of Technology, Pasadena, CA 91101, USA
\aff{3} Department of Mechanical Engineering, Massachusetts Institute of Technology, Cambridge, MA 02138, USA
\aff{4}D\'epartement Fluides, Thermique, Combustion, Institut PPrime, CNRS - Universit\'e de Poitiers - ENSMA, 86360, Chasseneuil-du-Poitou, France
\aff{5}Cascade R\&D, Cadence Design Systems, San Jose, CA 95134, USA}

\begin{document}
\maketitle

\nolinenumbers

\begin{abstract}

We study coherent structures in subsonic turbulent jets subject to a flight stream. A thorough characterisation of the effects of a flight stream on the turbulent field was recently performed by \citet{Maia_etal_flight_2023} and fluctuation energy attenuations were observed over a broad range of frequencies and azimuthal wavenumbers. The Kelvin-Helmholtz, Orr and lift-up mecahnisms were all shown to be weakened by the flight stream. Here we expand upon that study and model the changes in the dynamics of jets in flight using global resolvent analysis. The resolvent model is found to correctly capture the main effects of the flight stream on the dynamics of coherent structures, which are educed from a large-eddy simulation database using spectral proper orthogonal decomposition (SPOD). Three modifications of note are: the damping of low-frequency streaky/Orr structures that carry most of the fluctuation energy; a degradation of the low-rank behaviour of the jet in frequencies where modal instability mechanisms are dominant; and a rank decrease at very low Strouhal numbers. The latter effect is underpinned by larger gain separations predicted by the resolvent analysis, due to a reduction in the wavelength of associated flow structures. This leads to a clearer relative dominance of streaky structures generated by the lift-up mechanism, despite the fact that the lift-up mechanism has been weakened with respect to the static jet.

\end{abstract}



\section{Introduction}
\label{sec:intro}

The effect of forward flight on jet aeroacoustics has been a matter of industrial and scientific interest for decades. On the one hand, characterising and quantifying the acoustic field of jets in the presence of a flight stream is important for the certification of aircraft flyover noise \citep{VishwanathanFlight}. On the other hand, understanding how the flight stream modifies the acoustic field can help improve sound-source models \citep{Crighton_etal1977}. Numerous experimental studies have been performed in order to understand sound radiation from jets in flight \citep{VonGlahn, Cocking, Bushel, Packman, Plumbee, Bryce1984, VishwanathanFlight}. More recently, large-eddy simulation (LES) has been used to investigate the acoustics of jets with flight stream in free \citep{Zhong_etal_AIAA2017} and installed \citep{James_etal_AIAA2018} configurations.

Where the turbulent jet is concerned, \cite{TannaMorris1977} studied the effects of forward flight on flow statistics and observed that the flight stream modifies the development of the mean flow, producing a stretching of the potential core, and a reduction of both shear-layer thickness and turbulent kinetic energy. The latter effect is underpins the reduction of radiated sound pressure levels. \cite{MuchalkeHermann1982} recognised that the mean-flow modification leads to a stabilising effect on the Kelvin-Helmholtz (KH) instability. This was recently confirmed by \cite{SoaresAIAA2020} using a stability model based on the Parabolised Stability Equations (PSE) for different flight stream velocities. The effect of the flight stream on coherent structures is of some interest given their now recognised importance for jet dynamics and sound radiation \citep{JordanColoniusReview, CavalieriatalAMR2019}. Motivated by this, \cite{Maia_etal_flight_2023} performed a characterisation of the effect of a flight stream on the frequency-azimuthal wavenumber spectrum, using time-resolved Particle-Image Velocimetry (PIV) and large-eddy simulation (LES) databases. The study showed that the reduction in fluctuating energy, observed in early experiments \citep{TannaMorris1977}, is distributed over a broad region of frequency-azimuthal wavenumber space and that the attenuation of coherent structures is associated with a weakening of Orr, Kelvin-Helmholtz and lift-up instability mechanisms. Streaky structures with azimuthal wavenumbers in the range $1\leqslant m \leqslant 4$, that carry most of the fluctuation energy downstream of the end of the jet potential core, are the most strongly impacted by the flight stream. Locally-parallel, linear mean-flow model was also found to predict the overall attenuation trend of linear mechanisms observed in the data. However, due to the limitations of the locally-parallel framework, some aspects of jet dynamics with the flight stream could not be correctly modelled. For instance, the local model is not equipped to predict the correct shape of the energy spectrum at a given position, because it does not take into account the upstream amplification of flow structures. Analysis of Orr structures was not straightforward, because the local model requires knowledge of their streamwise wavenumber, which is unknown \textit{a priori}. Furthermore, an interesting rank-decrease in jet dynamics in flight, observed at low Strouhal numbers ($St$) through spectral proper orthogonal decomposition (SPOD), could not be explained by the model. In this work, we perform a global resolvent analysis, to overcome these issues, and to assess the extent to which linear mean-flow analysis can be used to explain the impact of a flight stream on the organisation of a turbulent jet.

Resolvent analysis has been widely used to model the mechanisms underpinning coherent structures observed in laminar and turbulent flows. In the latter case, linearisation is performed about the mean flow. In \enquote{static} conditions, the resolvent framework has been used to model coherent structures in turbulent jets, where it has allowed a classification of these according to the underlying growth mechanisms. The Kelvin-Helmholtz wavepacket is underpinned by convective modal instability. Whereas Orr-like structures arise when multiple, convectively stable modes, forced by ambient turbulence, combine linearly to produce transient growth on account of their non-normality. These structures have been characterised by the studies of \cite{Garnaud, Jeunetal, TissotJFM2017, SchmidtetalJFM2018, LesshafftPFR2019}. More recently, resolvent analysis has revealed the existence of the lift-up mechanism in turbulent jets \citep{NogueiraJFM2019, PickeringJFM2020, wang_lesshafft_cavalieri_jordan_2021}. This mechanism, characterised by higher azimuthal wavenumbers than the Orr mechanism, can also be understood in the linear mean-flow framework as arising from a non-normal linear combination of forced, convectively stable modes. The cited studies thus provide an explanation for streak-like structures that have been observed in numerous previous studies \citep{BeckerMassaro, BrowandLaufer, YuleJFM1978, Dimotakis1983, Agui1988, JungJFM2004}. 

For certain frequencies, the turbulent jet exhibits a low-rank behaviour \citep{SchmidtetalJFM2018, LesshafftPFR2019}, where the dynamics are largely dominated by the leading forcing and response modes. In that case, resolvent response modes are generally in good agreement with coherent structures educed from measurement or simulation data. When the dynamics are not low rank, there is no clear distinction between the leading and sub-optimal modes, and the leading response modes tend do differ substantially from empirical coherent structures. In that case, the nonlinear Reynolds stresses, treated as an endogenous forcing term in the resolvent framework, must be considered in order to achieve a complete picture of the coherent-structure dynamics. But that term is experimentally inaccessible, and even in high-fidelity simulations,its eduction is a delicate task \citep{Karban_etal_TCFD2022}.

In an attempt to improve the agreement between resolvent response modes and observed coherent structures, many recent studies have considered eddy-viscosity models. An eddy-viscosity can partially account for those non-linear effects of turbulence that attenuate the growth of coherent structures via a gradient-diffusion-like sink mechanism. Such models have been used for a variety of flows, both in the framework of stability \citep{Crouch_etal_JcP2007, Oberleithner_etal_JFM2014, Rukes_etal_2016,Kuhn_etal_JFM2021, Tammisola_Juniper_2016} and resolvent analysis \citep{HwangCossu2010, MorraetalArxiv2019, towne_lozano-duran_yang_2020, Pickeringetal_eddy_2021, Heidt_etal_AIAA2021}. A caveat of these models is that the modified forcing term loses its physical interpretability as the frequency-dependent Reynolds stresses and the resulting system of equations is no longer exact. However, the improved agreement between resolvent and SPOD modes that is observed when an eddy viscosity model is used \citep{Pickeringetal_eddy_2021} makes them appealing from the point of providing a basis suitable for description of empirical coherent structures and that would allow these and the mechanisms that drive them to be better understood.

The main contribution of the present work is the study of subsonic turbulent jets in the presence of a flight stream through global resolvent analysis. We analyse how coherent structures associated with different linear mean-flow mechanisms (characterised in static conditions by the studies cited above) are modified by the flight stream. While we focus here on the effect of the flight stream on the \textit{turbulent} field, the results can be used to inform sound-source models, such as those developed by \cite{Karban_etal2022, MaiaPRSA2019, CavalieriJitterJSV}. The remainder of the paper is organised as follows. In \S \ref{sec:les_setup}, we present the numerical databases used for the study. In \S \ref{sec:tools}, we describe the SPOD and resolvent frameworks used to educe and model coherent structures in the jet, respectively. In \S \ref{sec:en_maps} we present modal energy and amplification maps, and discuss how they are modified by the flight stream. This is followed in \S \ref{sec:coherent_structures} by a detailed analysis of coherent structures at different regions of the frequency-wavenumber plane. Finally, in \S \ref{sec:conclusions} we summarise the main conclusions of the study.

\section{Numerical database}
\label{sec:les_setup}

We explore two high-fidelity LES databases of subsonics jets at Mach number $M_j=0.9$ with and without flight streams. The simulations were performed using the compressible flow solver \enquote{CharLES} \citep{BresAIAA2017}, developed at Cascade Technologies, now part of Cadence Design Systems. Results for the case without the flight stream, $M_f=0$, were initially reported by \cite{BresJFM2018}. The present database are extensions of that study for both $M_f = 0$ and $0.15$ with longer databases and higher sampling frequency. All the large eddy simulations feature localized adaptive mesh refinement, synthetic turbulence and wall modeling on the internal nozzle surface (and external nozzle surfaces at $M_f = 0.15$) to match the fully turbulent nozzle-exit boundary layers in the experiments. The LES methodologies, numerical setup and comparisons with measurements are described in more details in \cite{BresJFM2018} and \cite{Maia_etal_flight_2023}.

The nozzle pressure ratio and nozzle temperature ratio are $NPR = P_t/P_f = 1.7$ and $NTR = T_t/T_f = 1.15$, respectively, and match the experimental conditions. The jet is isothermal ($T_j/T_f=1.0$), and the jet Mach number is $M_j = U_j/c = 0.9$. The subscript $_t$ refers to total conditions, $_j$ refers to jet exit conditions, and $_f$ to the flight stream. For both experiment and simulation, the Reynolds number is $Re = U_j D/\nu_j \approx 1\times10^6$, where $\nu$ is the kinematic viscosity, $U_j$ is the jet exit velocity and $D$ is the nozzle diameter, which is 50mm. Synthetic turbulence boundary conditions are applied inside and outside the nozzle surfaces to model the boundary layer trip used in the experiments 3D upstream of the nozzle exit. Simulation parameters and LES settings are shown in Table~\ref{table:LEScase}. To facilitate postprocessing and analysis, the LES data is interpolated from the original unstructured LES grid onto structured cylindrical grids in the jet plume and in the nozzle pipe. These structured cylindrical grids were originally designed for the grid with 16M control volumes, such that the resolution approximately corresponds to the underlying LES resolution. For both structured grids, the points are equally-spaced in the azimuthal direction to enable a Fourier series decomposition in azimuth. A detailed validation of the databases was carried out by \cite{Maia_etal_flight_2023} through comparisons with extensive PIV experiments.

\begin{table}
\begin{center}
\begin{tabular}{ l  c  c  c  c  c  c  c  c  }
       Case   name   &  Mesh size & $M_j$ & $M_f$ & $T_j/T_f$  & $Re$ & $dt c/ D$&  $ \Delta t  c /D$& $t_{sim} c /D$\\
    \textit {BL16M\_M09}           & $15.9\times10^6$   & 0.9  & 0  &     1.0&  $1 \times 10^6$ &      0.001 & 0.1 &3000   \\
    \textit{BL22M\_M09\_Mf015}        & $21.8\times10^6$  &  0.9 &0.15   &  1.0&  $1\times 10^6$ &0.001 & 0.1 &2000   \\

\end{tabular} 
\caption{Operating conditions and simulation parameters of the main LES, where $t_{sim}$ is the simulation time and $\Delta t$ is the sampling period of the database recording.}
\label{table:LEScase}
\end{center}
\end{table}

\section{Tools}
\label{sec:tools}

\subsection{Spectral Proper Orthogonal Decomposition}
\label{sec:spod}

SPOD is now a widely used tool for the of study turbulent flows. It decomposes the data into an orthogonal basis ranked in terms of an energy norm, and can provide a useful basis for the description of empirical coherent structures, particularly when the leading eigenvalue is substantially larger than its subdominante counterparts.

In the framework of SPOD, given the state vector, $\mathbf{q}=[\rho, u_{x},u_{r}, u_{\theta}, T]^{T}$, subject to a Reynolds decomposition,

\begin{equation}
\mathbf{q}(x,r,\theta,t) = \overline{\mathbf{q}}(x,r,\theta) + \mathbf{q}'(x,r,\theta,t),
\end{equation}
optimal modes for a given azimuthal wavenumber and Strouhal number pair, $\mathbf{\Psi}_{m,\omega}$ are obtained through eigendecomposition of the cross-spectral density (CSD) matrix, $\hat{\mathbf{S}}_{m,\omega}$,

\begin{equation}
\hat{\mathbf{S}}_{m,\omega}\mathbf{W}\mathbf{\Psi}_{m,\omega}=\mathbf{\Psi}_{m,\omega}\mathbf{\Lambda}_{m,\omega}.
\label{eq:spod}
\end{equation}
The cross-spectral density matrix is computed as $\hat{\mathbf{S}}_{m,\omega}=\hat{\mathbf{Q}}_{m,\omega}\hat{\mathbf{Q}}_{m,\omega}^{*}$, where $\hat{\mathbf{Q}}_{m,\omega}=[\hat{\mathbf{q}}_{m,\omega}^{(1)} \hat{\mathbf{q}}_{m,\omega}^{(2)} \cdots \hat{\mathbf{q}}_{m,\omega}^{(N_{blk})}]$ is the ensemble of $N_{blk}$ flow realisations at $(m,\omega)$, with $\hat{\mathbf{q}}_{m,\omega_{k}}^{(l)}$ denoting the $l$th Fourier realisation of the turbulent fluctuations, $\mathbf{q}'$, in time and azimuthal direction at the frequency $\omega$ and wavenumber $m$. The asterisk denotes conjugate transpose.

The eigenvalues, $[\lambda_{m,\omega}^{(1)}, \lambda_{m,\omega}^{(2)} \cdots \lambda_{m,\omega}^{(nblk)}]$ corresponding to the modal energy are organised in decreasing order in the diagonal matrix $\mathbf{\Lambda}_{m,\omega}$. The modes so obtained are orthogonal with respect to a given inner product,

\begin{equation}
\left< \mathbf{q}_{1},\mathbf{q}_{2}\right>=\mathbf{q}_{1}^{*} \mathbf{W} \mathbf{q}_{2}.
\label{eq:inner_product}
\end{equation}
Here we consider a weighting matrix, $\mathbf{W}$, describing Chu's compressible energy norm \citep{Chu1965},

\begin{equation}
\left< \mathbf{q}_{1},\mathbf{q}_{2}\right>_{E}=\iiint\mathbf{q}_{1}^{*}\mathrm{diag} \left(\frac{\bar{T}}{\gamma \bar{\rho}M_{j}^2},\bar{\rho}, \bar{\rho}, \bar{\rho}, \frac{\bar{\rho}}{\gamma(\gamma-1)\bar{T}M_{j}^2} \right)\mathbf{q}_{2} r\mathrm{d}x\mathrm{d}r \mathrm{d}\theta.
\label{eq:chu}
\end{equation}
The CSDs are computed using Welch's periodogram method. The data was segmented into blocks of 512 samples with 75\% overlap, resulting in a frequency resolution of $\Delta St=0.0217$.

\subsection{Resolvent Analysis}
\label{sec:model}
Since the works of \citet{HwangCossu2010} and \citet{MckeonSharma}, resolvent analysis has been extensively used to identify optimal forcing and response mechanisms in laminar and turbulent flows and to model coherent structures.
The analysis starts with the linearised Navier-Stokes equations in frequency domain, expressed in input-output form \citep{SchmidtetalJFM2018},

\begin{equation}
\left(i\omega\mathbf{I}-\mathbf{A}_m\right)\hat{\mathbf{q}}_{m,\omega} = \mathbf{B}\hat{\mathbf{f}}_{m,\omega},
\label{lin_ns1}
\end{equation}
\begin{equation}
\hat{\mathbf{y}}_{m,\omega} = \mathbf{C}\hat{\mathbf{q}}_{m,\omega},
\label{lin_ns2}
\end{equation}
where $\mathbf{A}_m$ is the linearised Navier-Stokes operator, $\hat{\mathbf{q}}_{m,\omega}$ is the Fourier-transformed state vector, and $\hat{\mathbf{f}}_{m,\omega}$ is a term representing the nonlinear Reynolds stresses, which are treated as an endogenous forcing term. The subscript $_{m,\omega}$, with $m$ the azimuthal wavenumber and $\omega$ the frequency, denotes Fourier transform in the azimuthal and time directions. $\hat{\mathbf{y}}_{m,\omega}$ defines the derired response, or output, as a function of the state. $\mathbf{I}$ is the identity matrix, and $\mathbf{B}$ and $\mathbf{C}$ are matrices that can be used to restrict forcing and observation to specific regions of space and/or to a limited number of forcing and response terms.

Input and output are related through,

\begin{equation}
\hat{\mathbf{y}}_{m,\omega} = \mathbf{R}_{m,\omega}\hat{\mathbf{f}}_{m,\omega},
\end{equation}
where $\mathbf{R}_{m,\omega}$ is the resolvent operator,

\begin{equation}
\mathbf{R}_{m,\omega} = \mathbf{C}\left(i\omega\mathbf{I}-\mathbf{A}_m\right)^{-1}\mathbf{B}.
\end{equation}
We then define a weighted resolvent operator, $\tilde{\mathbf{R}}_{m,\omega}$ by introducing Chu's compressible energy norm through the matrix $\mathbf{W}$,

\begin{equation}
\tilde{\mathbf{R}}_{m,\omega} = \mathbf{W}^{1/2}\mathbf{R}_{m,\omega}\mathbf{W}^{-1/2}.
\end{equation}

The goal of resolvent analysis is to seek an optimal forcing that maximises the norm of the associated flow response,

\begin{equation}
\sigma_1^2 = \max_{||\hat{\mathrm{f}}_{m,\omega}||=1}\frac{\left|\left| \tilde{\mathbf{R}}_{m,\omega}\hat{\mathbf{f}}_{m,\omega} \right|\right|^{2}_{\mathbf{W}}}{\left|\left|\hat{\mathbf{f}}_{m,\omega} \right|\right|^{2}_{\mathbf{W}}}.
\end{equation}
This can be achieved through singular-value decomposition (SVD) of the resolvent operator,

\begin{equation}
\tilde{\mathbf{R}}_{m,\omega} = \mathbf{U}\mathbf{\Sigma}\mathbf{V}^{*}.
\end{equation}
Forcing ($\mathbf{u}_{i}$) and response ($\mathbf{v}_{i}$) modes are defined as $\mathbf{v}_{i} = (\mathbf{W}^{-1/2})^{*}\mathbf{V}_{i}$ and $\mathbf{u}_{i} = (\mathbf{W}^{-1/2})^{*}\mathbf{U}_{i}$, with $i$ denoting the $i$-th column of $\mathbf{V}$ and $\mathbf{U}$. The singular values associated with each forcing-response pair, $\sigma_{i}$, are arranged in descending order in the diagonal matrix $\mathbf{\Sigma}$. Optimal forcings and responses for each $m, \omega$ pair are then given by $\mathbf{v}_{i}$ and $\mathbf{u}_{1}$, and their associated energy gain is the square of the leading singular value, $\sigma_1^2$. As in \cite{SchmidtetalJFM2018}, the governing equations are discretized using fourth-order summation by parts finite differences \citep{mattsson2004summation}, the polar singularity is treated as in \cite{mohseni2000numerical}, and non-reflecting boundary conditions/sponges are employed at the domain boundaries. The dominant resolvent modes are computed using randomized linear algebra methods \citep{martinsson2019randomized} that allow for the efficient computation of dominant singular values/vectors.

As mentioned above, we focus on characterisation of the flight stream effect on the turbulent field, rather than the acoustic field, as done by many previous studies \citep{VonGlahn, Cocking, Bushel, Packman, Plumbee, Bryce1984, MorfeyTester, VishwanathanFlight}. To that end, we restrict the forcing and observation fields to the hydrodynamic region via the matrices $\mathbf{B}$ and $\mathbf{C}$. This is done by setting the elements of those matrices to one within the jet shear-layer, delimited by the region where $\overline{U}_x/\overline{U}_j \geqslant 0.05$, and gradually reducing them to zero for larger $r$. Here we also enforce a mask on the forcing field inside the jet potential core. The reason for this is that the jet supports trapped acoustic waves that resonate, producing a tonal dynamics in the potential core \citep{schmidt_etal_2017, towne_etal_2017}. The resonant mechanism at $M_j=0.9$ is such that the jet is marginally globally stable at certain frequencies, and the resolvent analysis thus identifies this resonance mechanism as a leading candidate for optimal growth. As this mechanism is not dominant in the SPOD analysis, as we will briefly show, we choose to mask it, so as to focus on the dominant turbulent structures. In appendix \ref{appA} we briefly discuss their presence in the forcing and response modes, and we show that their suppression does not affect the analysis of other mechanisms.

\begin{figure}
\centering
\includegraphics[trim=12cm 8cm 12cm 7.5cm, clip=true,width=0.9\linewidth]{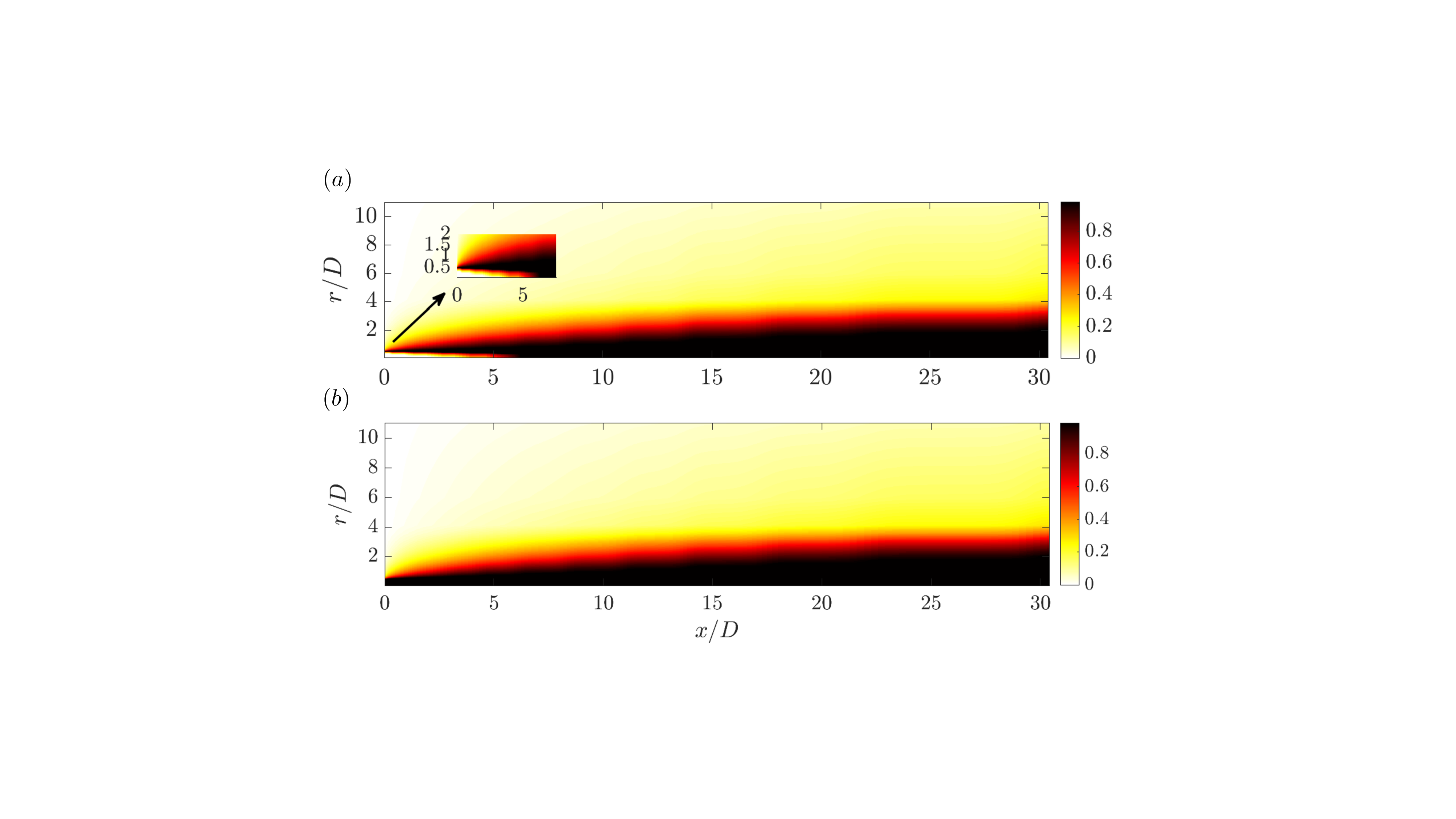}
\caption{Example of mean-flow masks applied to (a) the forcing and (b) the response for the static case, $M_f=0$.}
\label{fig1}
\end{figure}

Figure \ref{fig1} shows the resulting mean-flow masks applied to the forcing and response fields. The goal of the resolvent analysis is therefore to seek an optimal forcing that maximises the compressible energy norm of the associated flow response in the highlighted region.
 
\subsubsection{Eddy-viscosity model}

The effects of eddy-viscosity on the predictive capabilities of linear analysis have been explored in a number of recent studies \citep{SchmidtetalJFM2018, Crouch_etal_JcP2007, Oberleithner_etal_JFM2014, Rukes_etal_2016, Tammisola_Juniper_2016, SchmidtetalJFM2018, HwangCossu2010, MorraetalArxiv2019, Pickeringetal_eddy_2021, Kuhn_etal_JFM2021}. The study of \cite{Pickeringetal_eddy_2021} showed substantial improvements in the agreement between resolvent response modes and coherent structures educed from flow data through SPOD.

Here we adopt the mean-flow-consistent eddy viscosity model of \cite{Pickeringetal_eddy_2021}. The molecular viscosity, $\mu$, is replaced by the sum $\mu+ \mu_T$ in the linearised equations, and a suitable spatial structure for the eddy viscosity, $\mu_T$, is determined, as described below. The resolvent operator can then be rewritten as

\begin{equation}
\mathbf{R}_{m,\omega} = \mathbf{C}\left(i\omega\mathbf{I}-\mathbf{A}_m - \mathbf{A}_{m,T}(\mu_T) \right)^{-1}\mathbf{B},
\end{equation}  
where $\mathbf{A}_{m,T}$ only possess terms including $\mu_T$. The reader is referred to \cite{Pickeringetal_eddy_2021} for the equations for the modified operator.

The eddy-viscosity field, $\mu_T(x,r)$, is found through an optmisation procedure that minimises the error by which the mean flow satisfies the zero-frequency, zero-azimuthal wavenumber linearised Navier-Stokes equations, modified with the addition of an eddy-viscosity field. Details about the method can be found in \cite{Pickeringetal_eddy_2021}. 

\begin{figure}
\centering
\includegraphics[trim=12cm 8cm 12cm 7.5cm, clip=true,width=0.9\linewidth]{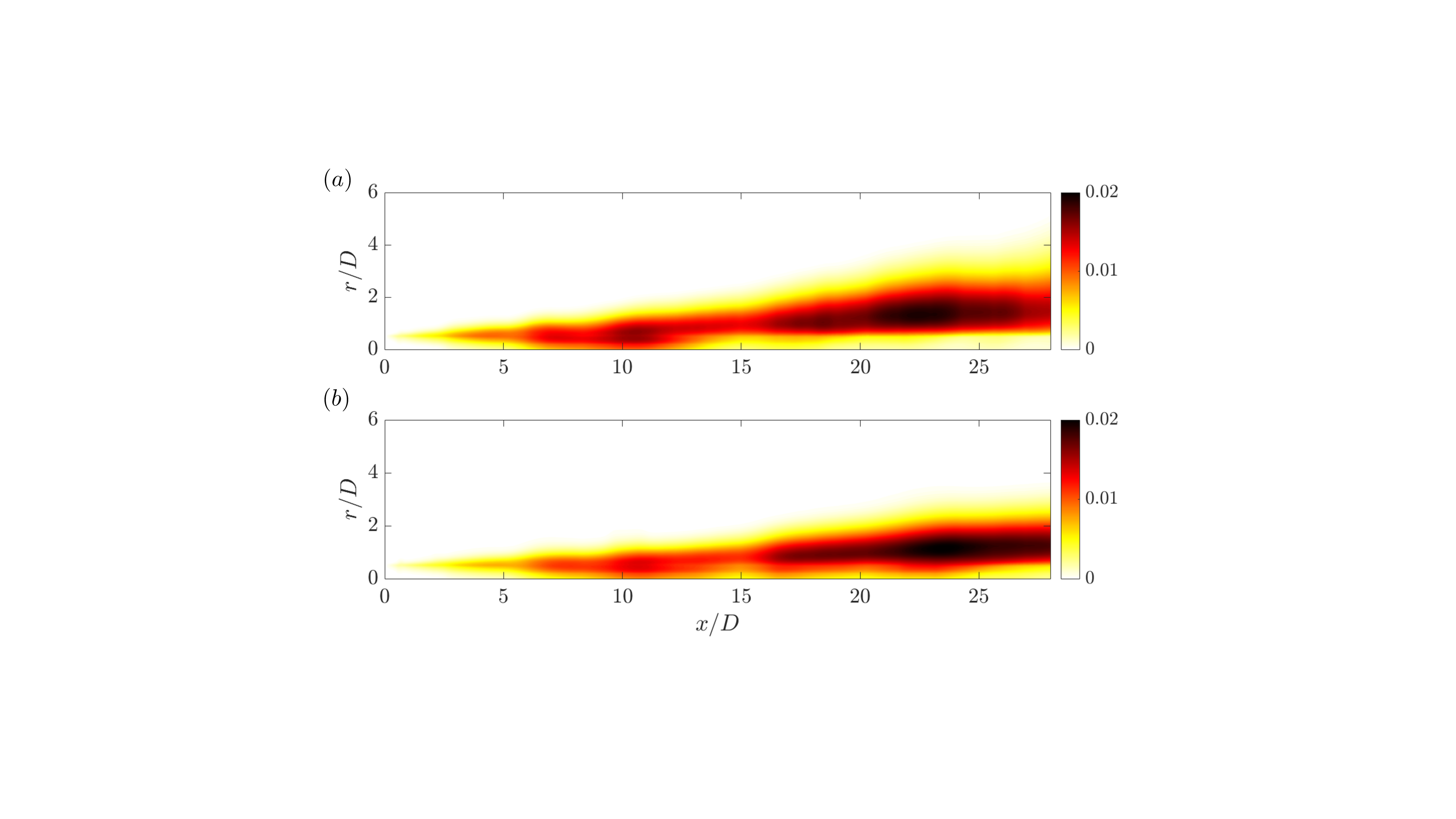}
\caption{Mean-flow-consistent eddy-viscosity fields computed at zero frequency and azimuthal wavenumber for (a) the static case, $M_f=0$, and (b) the flight stream case, $M_f=0.15$.}
\label{fig2}
\end{figure}

Figure \ref{fig2} shows the mean-flow-consistent eddy-viscosity fields for the $M_f=0$ and $M_f=0.15$ jets. Their shapes are similar to those obtained at lower Mach number by \citep{Pickeringetal_eddy_2021}. The amplitude of the eddy-viscosity model was scaled by a constant, $c=0.15$. This value was selected based on the best alignment achieved with respect to leading SPOD modes at different frequencies and azimuthal wavenumbers.

\section{Modal energy and amplification maps}
\label{sec:en_maps}

Leading SPOD modes reveal coherent structures, as mentioned above, that can frequently be associated with linear mean-flow mechanisms in jets \citep{CavalieriatalAMR2019, NogueiraJFM2019, PickeringJFM2020}. We compare the modal energy of leading SPOD modes, $\lambda_1(\omega,m)$, and the optimal resolvent gains, $\sigma_1^2(\omega,k)$. The association of leading SPOD modes with a given mechanism can be made according to the regions of dominance of each mechanism in frequency-wavenumber space, as characterised in previous studies \citep{Garnaud, Jeunetal, TissotJFM2017, SchmidtetalJFM2018, LesshafftPFR2019, NogueiraJFM2019, PickeringJFM2020} (figure \ref{fig6} roughly delimits those regions, based on the work of \citet{PickeringJFM2020}).

\begin{figure}
\centering
\includegraphics[trim=0.5cm 4.5cm 0.5cm 4.5cm, clip=true,width=\linewidth]{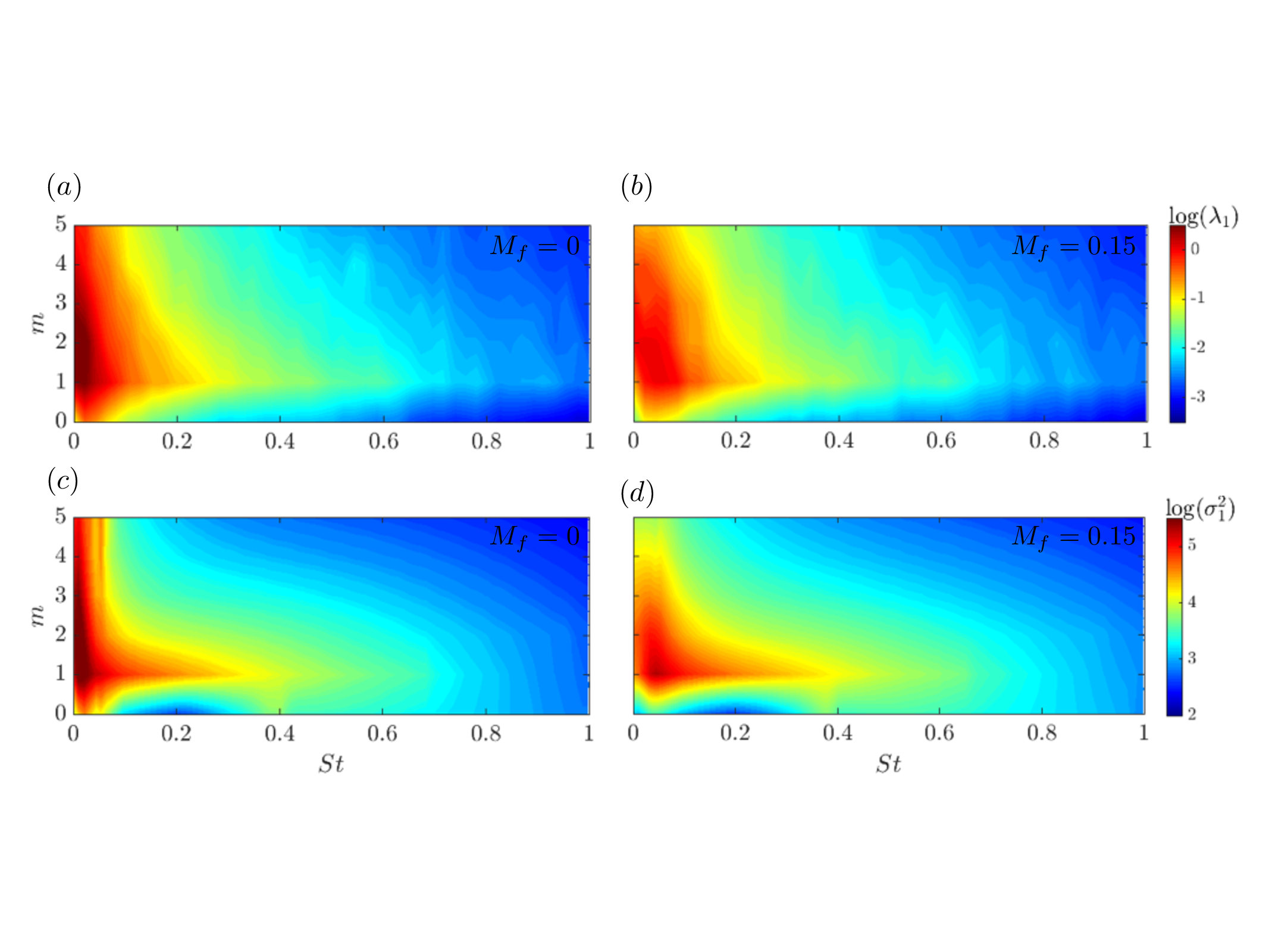}
\caption{Modal energy maps from SPOD and resolvent analysis. (a) and (b); modal energy of the leading SPOD mode, $\lambda_1$. (c) and (d): leading resolvent gain, $\sigma_1^2$. Contours are in $\mathrm{log}_{10}$ scale.}
\label{fig3}
\end{figure}

Figure \ref{fig3} shows maps of modal energy and leading resolvent gains in $St$-$m$ space for static, $M_f=0$, and flight, $M_f=0.15$ conditions. The modal energy maps reveal that most of the flow energy is in the $St \to 0$ zone of the spectrum, and is carried by streaky structures \citep{NogueiraJFM2019, PickeringJFM2020}. The map of the flight stream case shows a striking attenuation in that region of the spectrum, especially for azimuthal modes $m=1$-$4$. At higher $St$, mode $m=1$ is the most energetic, for both flow conditions. The resolvent gain spectrum is similar to the SPOD modal energy maps. The regions of high SPOD energy correspond, to a great extent, to the zones of maximum amplification predicted by the resolvent model. Modes $m=1$-$4$ possess the largest amplification rates in the $M_f=0$ case and are those most impacted by the flight stream, in agreement with what is observed in the flow data. This is a first indication that the resolvent analysis correctly captures the leading-order effect of flight on the turbulent kinetic energy.

\begin{figure}
\centering
\includegraphics[trim=2.5cm 1cm 3cm 2cm, clip=true,width=0.8\linewidth]{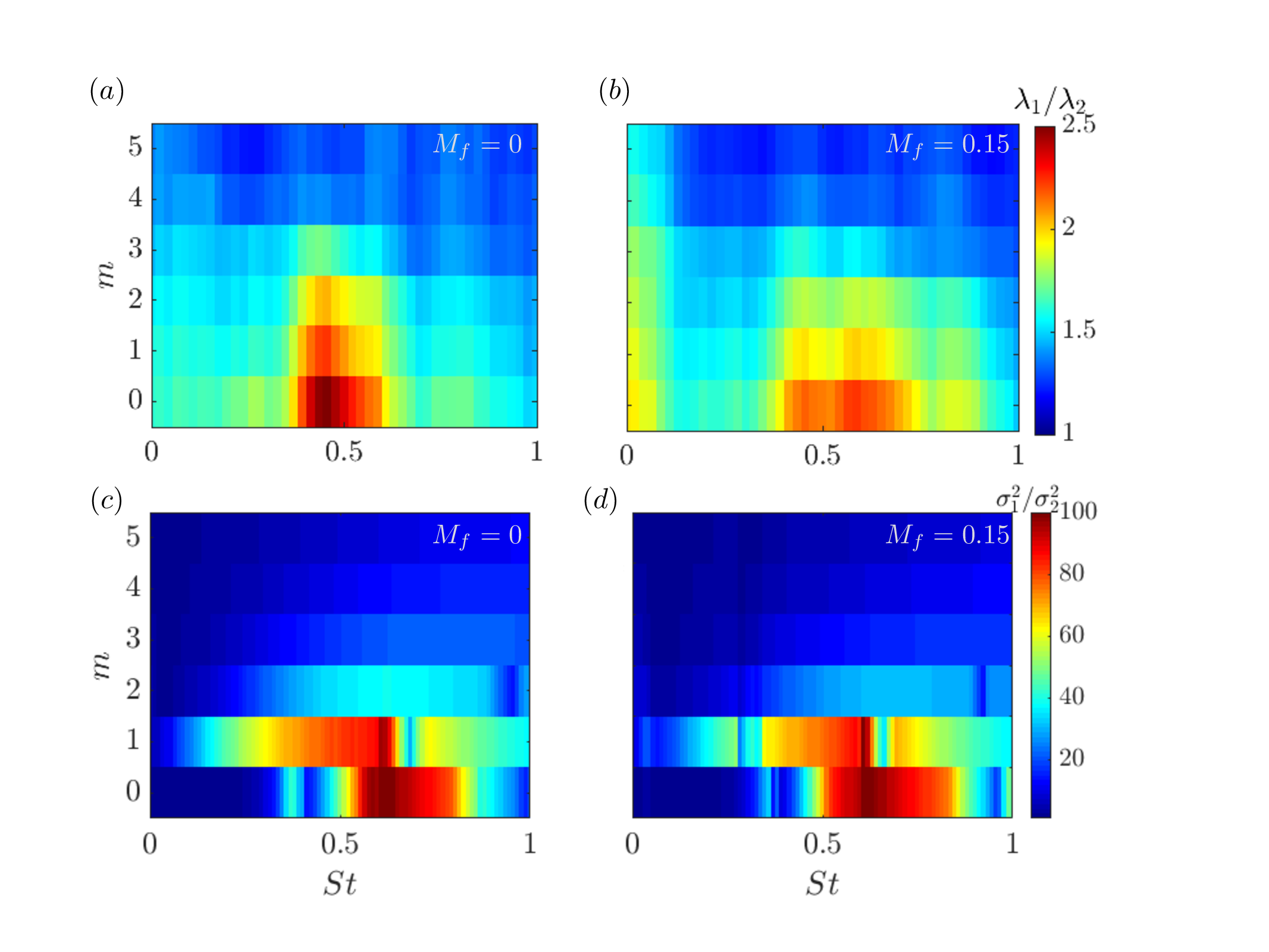}
\caption{Maps of eigenvalue separation , $\lambda_1/\lambda_2$, from SPOD (a-b) and gain separation, $\sigma_1^2/\sigma_2^2$, from resolvent analysis (c-d), highlighting zones of low-rank jet dynamics.}
\label{fig4}
\end{figure}

Figure \ref{fig4} shows contours of the ratio between the leading and second SPOD eigenvalues, $\lambda_1/\lambda_2$, and the ratio between the optimal and first suboptimal resolvent gains, $\sigma_1^2/\sigma_2^2$. The eigenvalue separation map for the $M_f=0$ case shows large peaks for the first three azimuthal wavenumbers around $St=0.4$-$0.5$, due to the KH modal instability mechanism. The flight stream reduces the peak values, but produces a broader region of low-rank behaviour, and a slight shift of that region towards higher $St$. These trends were shown by \cite{Maia_etal_flight_2023} to be consistent with a larger range of unstable frequencies in the flight case, and a shift of the most unstable KH mode towards higher frequency. The low-rank behaviour is found to be more pronounced in the resolvent model, including a large gain separation region for the $m=1$ mode at low $St$, which is not observed in the data. Unlike the SPOD maps, no weakening of the low-rank behaviour is observed in the resolvent model in the zone $0.3 \lessapprox St \lessapprox 1$, which is dominated by modal instability mechanisms for low azimuthal mode order. This is probably associated with the role of the nonlinear-forcing in the downstream region of the jet, the effect of which is present in the SPOD maps, but not in the resolvent maps. Furthermore, the KH instability mechanism is dominant in the initial jet region, up to the end of the potential core \citep{JordanColoniusReview}, and therefore its energy is inevitably masked in the global analysis by that of the most energetic, low-frequency structures that dominate the flow far downstream. In Appendix \ref{appB} we show results of resolvent analyses performed with domains truncated at $x/x_c=1$ and $x/x_c=1.5$, where $x_c$ is the potential core length, which is lenghtened by the flight stream. The truncations highlight the zone where KH wavepackets are convectively unstable. The results of the analysis show more pronounced effects of the flight stream on the amplification and gain separation associated with the modal instability mechanism.

\begin{figure}
\centering
\includegraphics[trim=3cm 8.5cm 3cm 6cm, clip=true,width=0.8\linewidth]{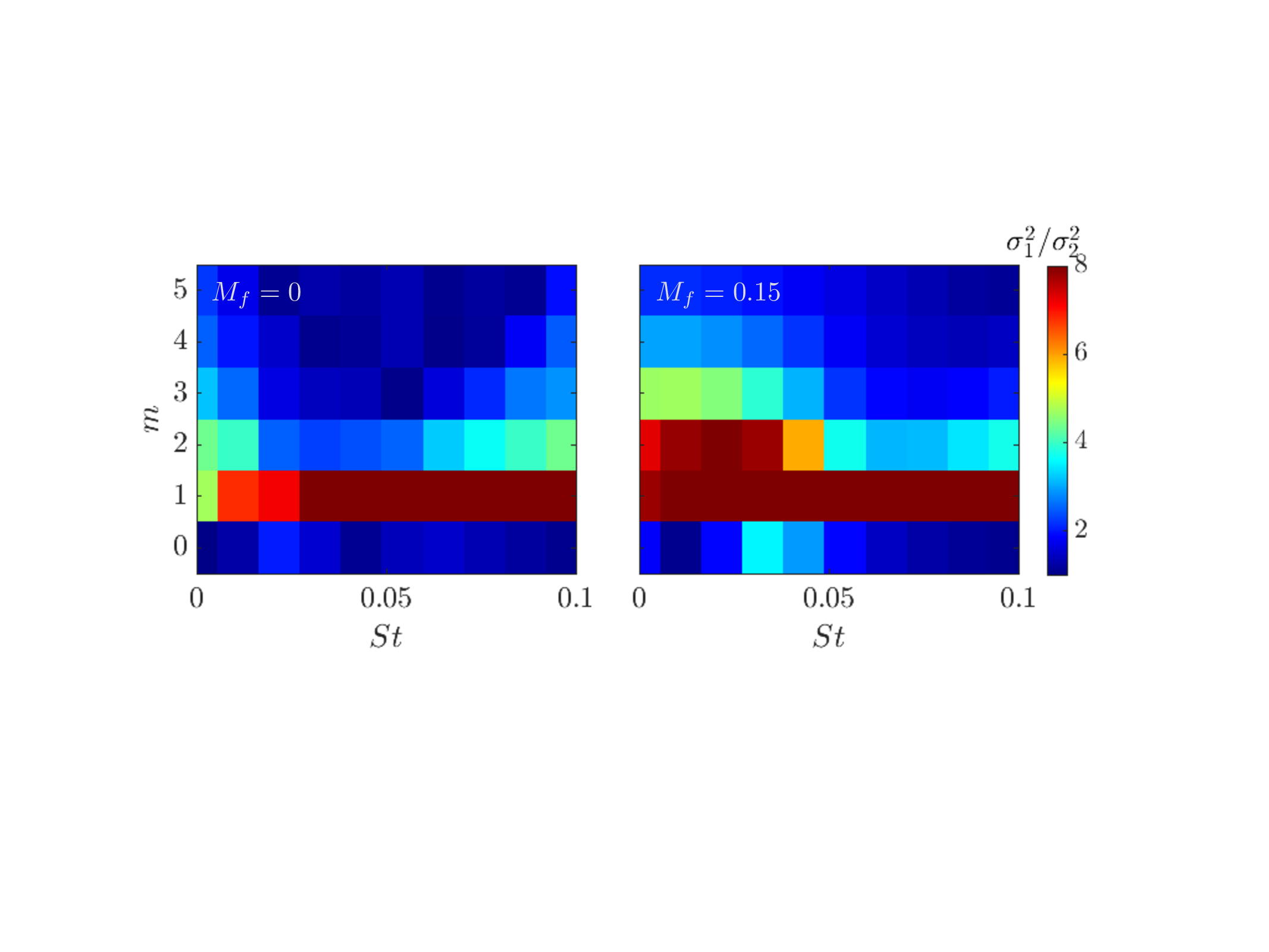}
\caption{Resolvent gain separation, $\sigma_1^2/\sigma_2^2$ close to the $St \to 0$ limit.}
\label{fig5}
\end{figure}

Interestingly, in the $St \to 0$ zone, the flight stream produces an increase in the the ratio $\lambda_1/\lambda_2$. This trend is found to be captured by the resolvent model. This can be seen in figure \ref{fig5}, which shows the low $St$ zone in more details. The $\sigma_1^2/\sigma_2^2$ is clearly enhanced for the helical modes in that zone as a result of the mean-flow modification in the flight stream case. An interpretation for this behaviour will be provided in the following sections.

\section{Mode shapes}
\label{sec:coherent_structures}

\begin{figure}
\centering
\includegraphics[trim=2.5cm 8cm 3cm 7cm, clip=true,width=\linewidth]{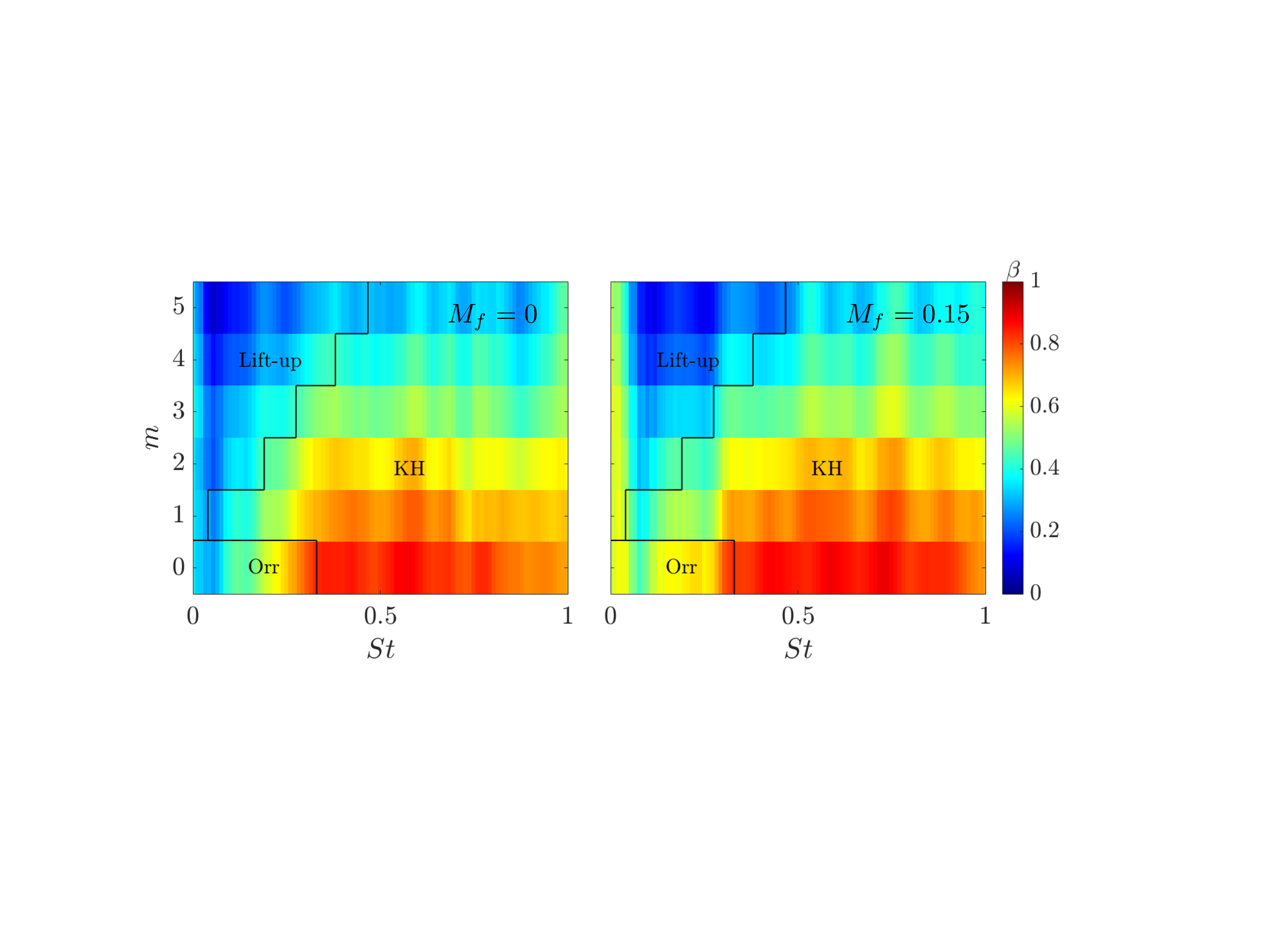}
\caption{Maps of alignment, measured by the $\beta$ metric, between leading SPOD and resolvent modes. The black lines are from \cite{PickeringJFM2020}, and approximately delimit the regions of dominance of the KH, Orr and lift-up mechanisms.}
\label{fig6}
\end{figure}

In this section, we perform a detailed comparison of coherent structures, educed through SPOD, and optimal resolvent modes in the static and flight cases. We define a projection coefficient,

\begin{equation}
\beta = \left| \mathbf{u}_{1}^{*} \mathbf{W} \mathbf{\Psi}_{1}\right|,
\end{equation}
that quantifies the alignment between SPOD and resolvent response modes. The metric varies between 0, in which case the modes are orthogonal to each other, to 1, for perfect alignment. Figure \ref{fig6} shows $\beta$ as a function of $St$ and $m$. The maps are divided in regions that roughly delimit the regions of dominance of the linear mechanisms studied. This is useful for the purpose of discussing each mechanism separately. But we emphasise that they are not intended to suggest that such a clear demarcation exists between the different mechanisms. 

Good alignment is obtained between SPOD and response modes in KH-dominated zones, for both flow conditions. A frequency shift is observed in the peak values of $\beta$ with the flight stream, following the changes in the modal stability characteristics discussed above. The alignment is noticeably worse in the $St<0.2$, in the Orr- and Lift-up-dominated zones. A similar trend was observed by \cite{Pickeringetal_eddy_2021}, even with optimal eddy-viscosity models, and this highlights the reliance of the non-modal mechanisms on the \enquote{source} component of endogenous forcing, which cannot be mimicked by a gradient-diffusion, eddy viscosity \enquote{sink}. Due to the small gain separation between in this region, suboptimal modes are as important as the leading mode, and the details of the nonlinear forcing projection on the input space are necessary for a correct description of the dynamics. We notice, however, an improved alignment in the $St \to 0$ region in the $M_f=0.15$ case. This region of improved alignment overlaps with the region where higher eigenvalue and gain separations were observed with the flight stream (figure \ref{fig4} and \ref{fig5}). The flight stream appears to create a more organised, low-frequency dynamics, despite the global weakening of linear, mean-flow growth mechanisms.

\begin{figure}
\centering
\includegraphics[trim=2.5cm 8cm 3cm 7cm, clip=true,width=\linewidth]{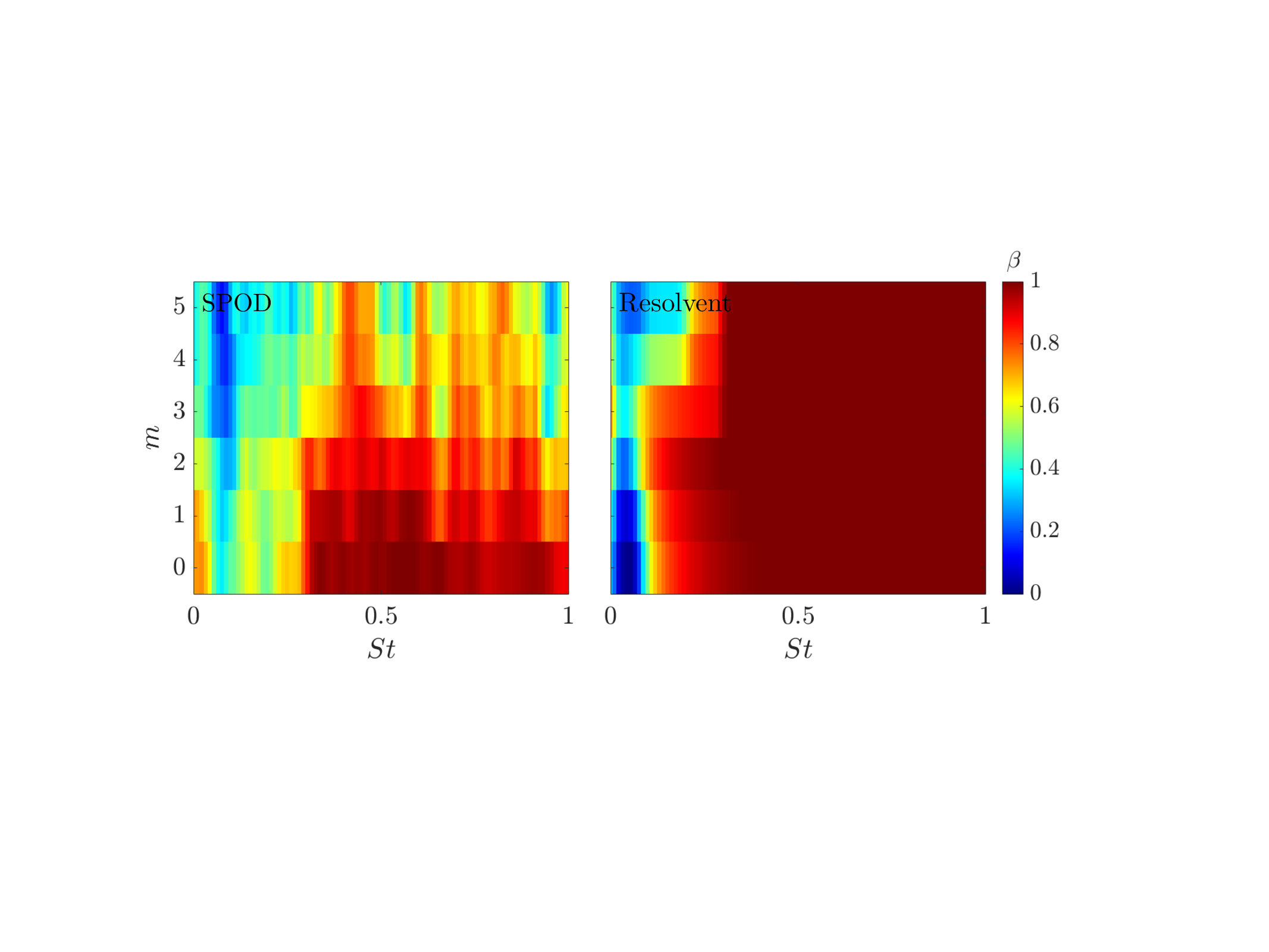}
\caption{Alignment between leading modes in the static and flight cases, measured by the $\beta$ metric. The left chart represents the alignment between SPOD modes and the right chart the alignment between resolvent modes.}
\label{fig7}
\end{figure}

\citet{Maia_etal_flight_2023} characterised the alignment between SPOD modes in the static and flight cases using the same metric defined above, but scaling the mean flows by the potential core length, $x_c$, which increases in the presence of the flight stream. It was shown that $\beta$ assumes high values in the KH-dominated zone, showing that the change in organisation of KH structures is largely dictated by the stretching of the potential core. The static-flight alignment in the Orr and lift-up dominated zones, on the other hand, was found to be much poorer. Here we extend that comparison to the leading resolvent modes in static and flight conditions, and compare it with SPOD results. The results are displayed in figure \ref{fig7}. The agreement between KH wavepackets in static and flight conditions is even more striking in the model; throughout the KH-dominated zone, the alignement is virtually perfect. As in the SPOD map, the alignment between resolvent modes deteriorates in the Orr and lift-up regions, showing that their reorganisation by the flight stream is more subtle than a simple mean-flow stretching.

In the following, we analyse separately the shapes of coherent structures associated with the three instability mechanisms with and without the flight stream. As mentioned above, previous studies have provided extensive characterisations of such structures in static conditions. Therefore, here we focus mainly on the modifications observed in flight condition. In what follows, the streamwise coordinate is scaled by the potential core length, as done by \cite{Maia_etal_flight_2023}.

\subsection{KH mechanism}

\begin{figure}
\centering
\includegraphics[trim=0cm 3cm 3cm 0cm, clip=true,width=0.85\linewidth]{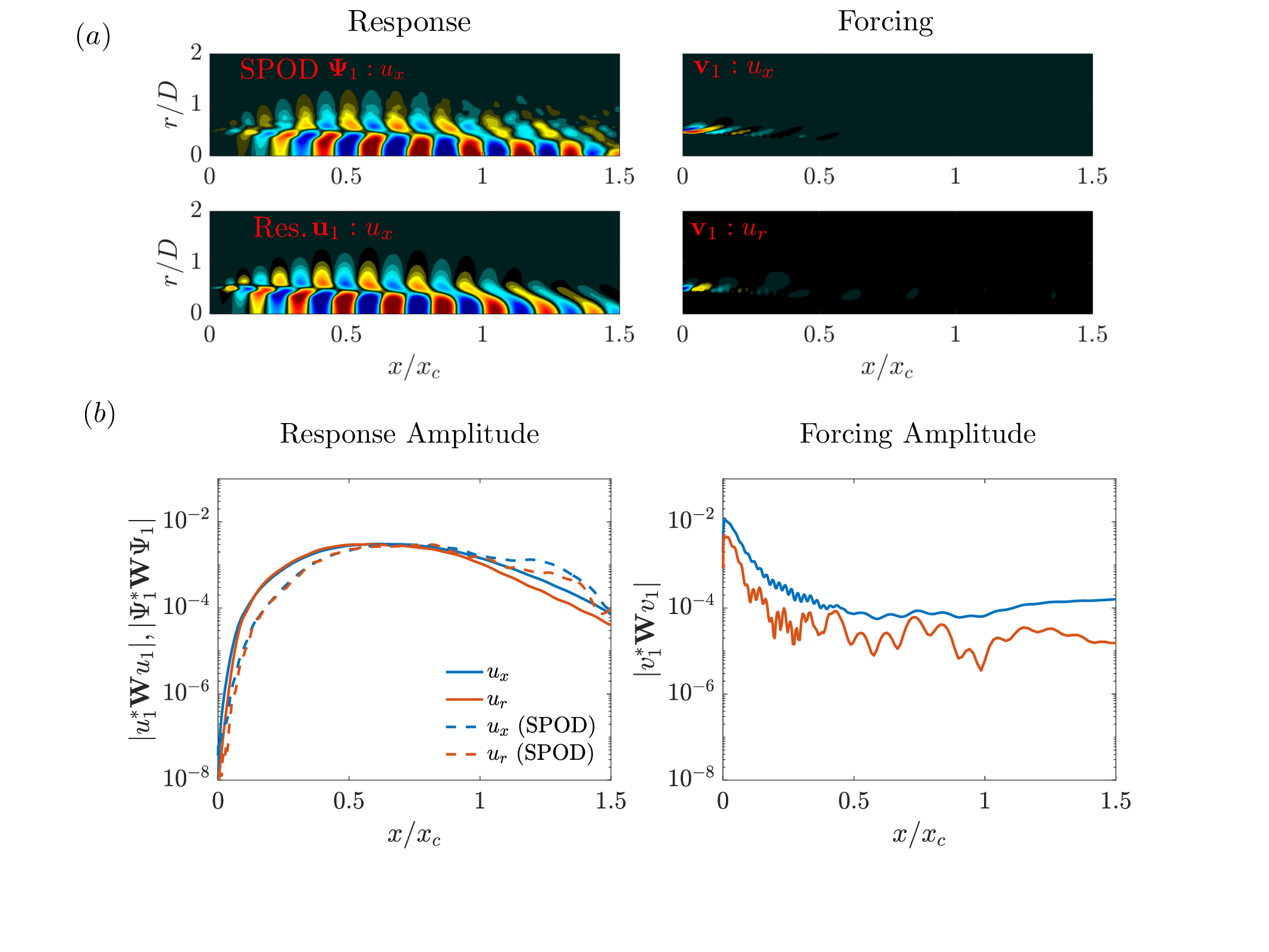}
\caption{(a) Leading forcing and response modes of the axisymmetric wavenumber $m=0$ and Strouhal number $St=0.6$ for the static case, $M_f=0$. Leading SPOD mode is also shown for comparison. The modes are shown with contours corresponding to $\pm 0.7||u_{x,r}||$. (b)  Component-wise amplitudes as a function of streamwise coordinate, computed through the compressible energy norm.}
\label{fig8}
\end{figure}

\begin{figure}
\centering
\includegraphics[trim=0cm 3cm 3cm 0cm, clip=true,width=0.85\linewidth]{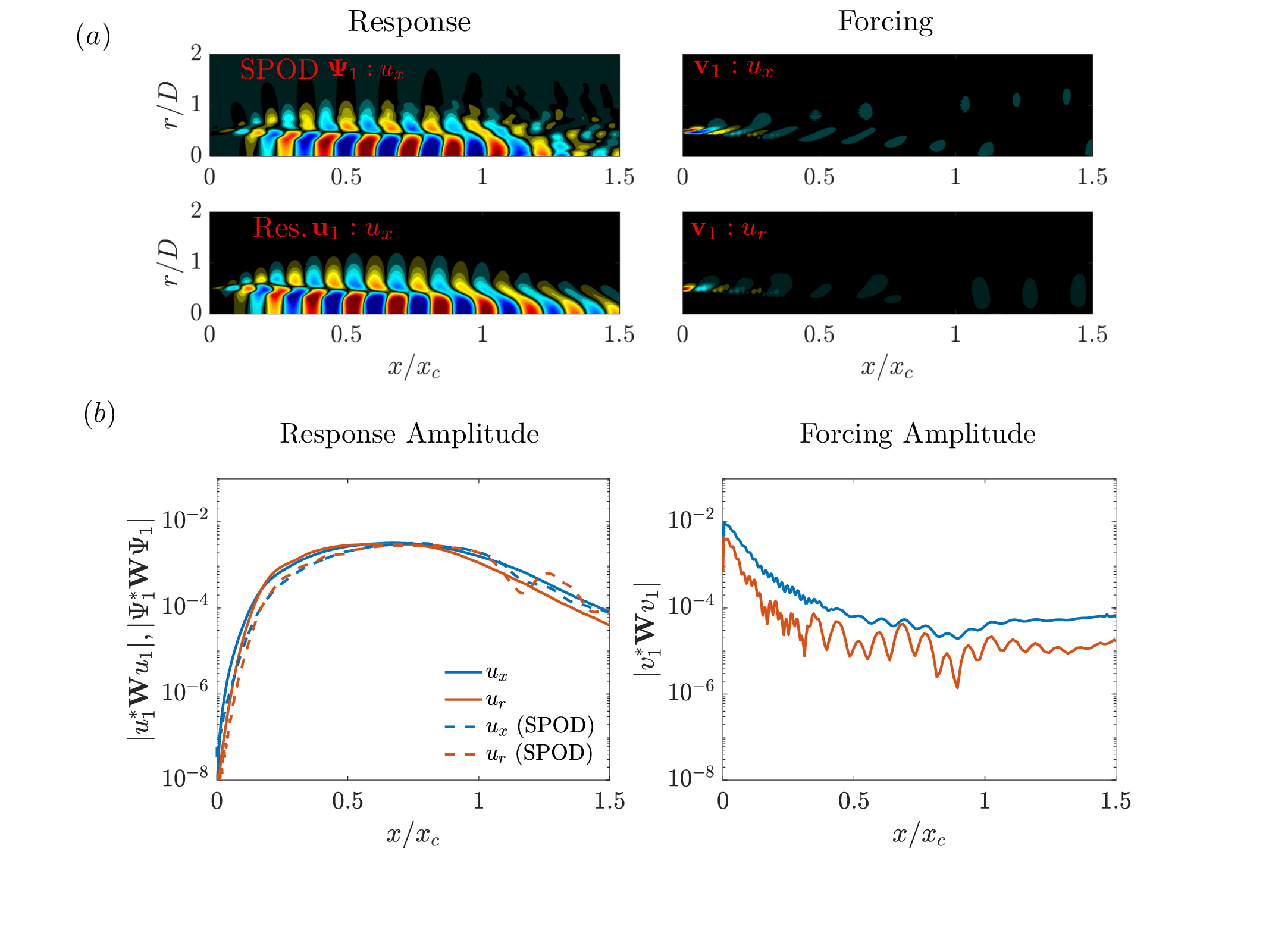}
\caption{Leading forcing and response modes of the axisymmetric wavenumber $m=0$ and Strouhal number $St=0.6$ for the flight case, $M_f=0.15$. Legend is as in figure \ref{fig8}.}
\label{fig9}
\end{figure}

The modal KH mechanism is dominant over a broad frequency range, $St \gtrsim 0.2$ \citep{SchmidtetalJFM2018, LesshafftPFR2019,PickeringJFM2020} and can be observed up to $St=4$ and $m=4$ near the nozzle region \citep{SasakiRapids}. Here we present results for the axisymmetric azimuthal mode at $St=0.6$ as a representative case where KH wavepackets are clear in the jet response. Similar trends were found for other azimuthal wavenumbers and Strouhal numbers within the KH-dominated region of the spectrum, indicated in figure \ref{fig6}. Figures \ref{fig8} and \ref{fig9} show leading forcing, $\mathbf{v}_1$, and response, $\mathbf{u}_1$ modes for the static and flight cases, respectively. The leading SPOD mode of streamwise velocity is also shown for comparison, and is in striking agreement with the leading response mode in both cases, consistent with the alignment metric shown in figure \ref{fig6}. The forcing modes exhibit Orr-like structures localised in the vinicity of the nozzle lip, and are in agreement with observations made at lower-Mach-number jets \citep{Garnaud, SchmidtetalJFM2018, LesshafftPFR2019}. Similar structures have also been observed within the nozzle boundary layer \citep{KaplanJFM2021}. The figures also show the streamwise evolution of the response and forcing amplitudes for each velocity component. The amplitudes are computed through the local compressible inner products, $|\mathbf{u}_1^{*}\mathbf{W}\mathbf{u}_1|$ and $|\mathbf{v}_1^{*}\mathbf{W}\mathbf{v}_1|$ for the resolvent, and $|\Psi_1^{*}\mathbf{W}\Psi_1|$ for the SPOD modes, at each streamwise position. Note that the $u_\theta$ component is null for the axisymmetric mode. The forcing amplitudes display a noisy behaviour in the initial jet region, as opposed to the smooth decay observed a lower Mach number \citep{PickeringJFM2020}. This behaviour is due to the signature of trapped waves. Despite the mask in the potential core being able to significantly attenuate these waves (see for instance the results of figure \ref{fig15} without the core mask), it does not eliminate them altogether. A more efficiently way to suppress them completely would be to also restrict the response at the jet core; but this would also impact the growth of KH waves in that zone and therefore it has not been done here. The stucture of KH wavepackets in static and flight conditions are found to be quite similar (which can also be inferred from the alignment maps of figure \ref{fig7}). With the potential core scaling, the regions of exponential, growth, stabilisation and decay are found to be quite similar.

\subsection{Orr mechanism}

\begin{figure}
\centering
\includegraphics[trim=0cm 3cm 3cm 0cm, clip=true,width=0.85\linewidth]{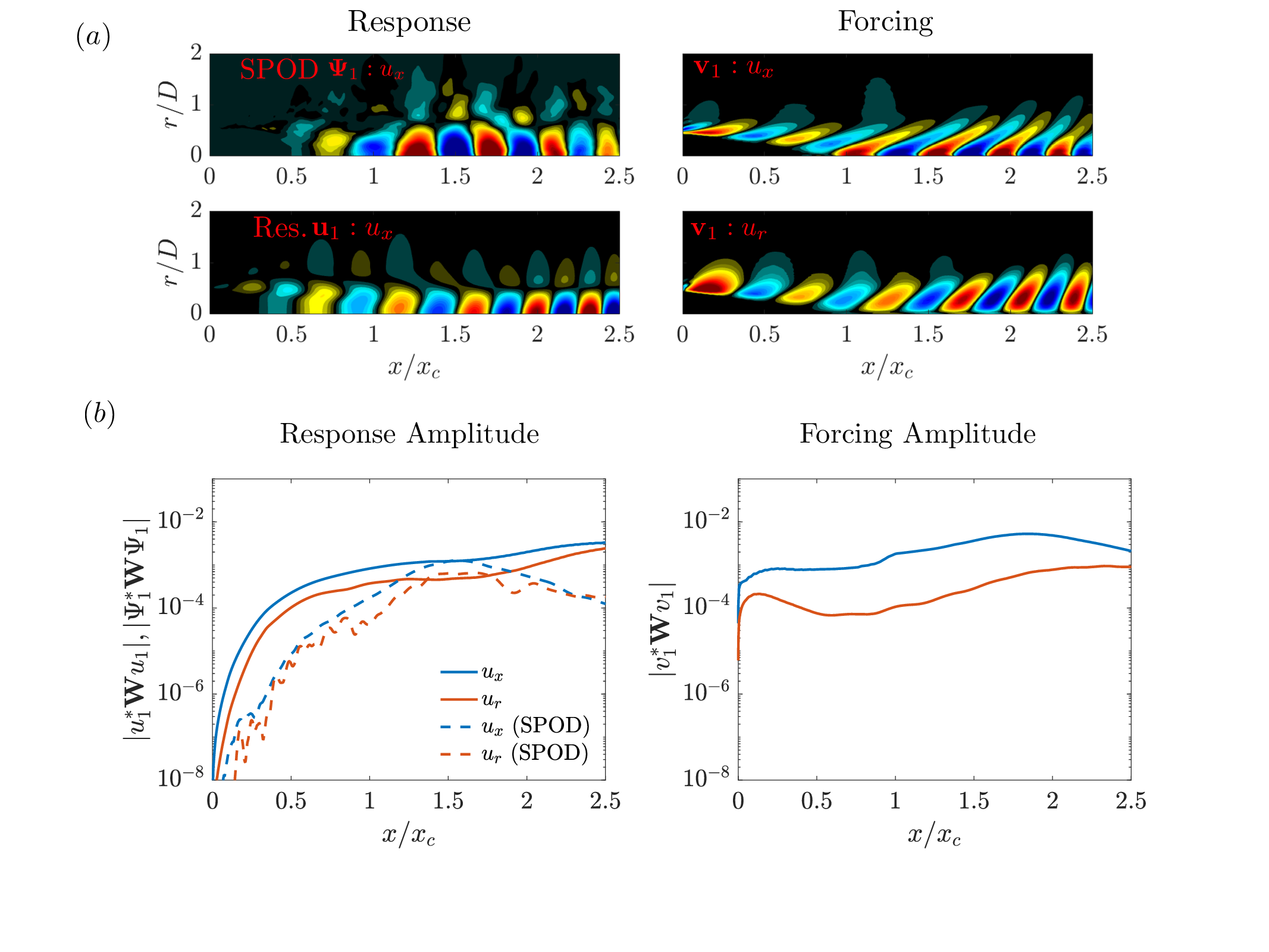}
\caption{Leading forcing and response modes of the axisymmetric wavenumber $m=0$ and Strouhal number $St=0.2$ for the static case, $M_f=0$. Legend is as in figure \ref{fig8}.}
\label{fig10}
\end{figure}

\begin{figure}
\centering
\includegraphics[trim=0cm 3cm 3cm 0cm, clip=true,width=0.85\linewidth]{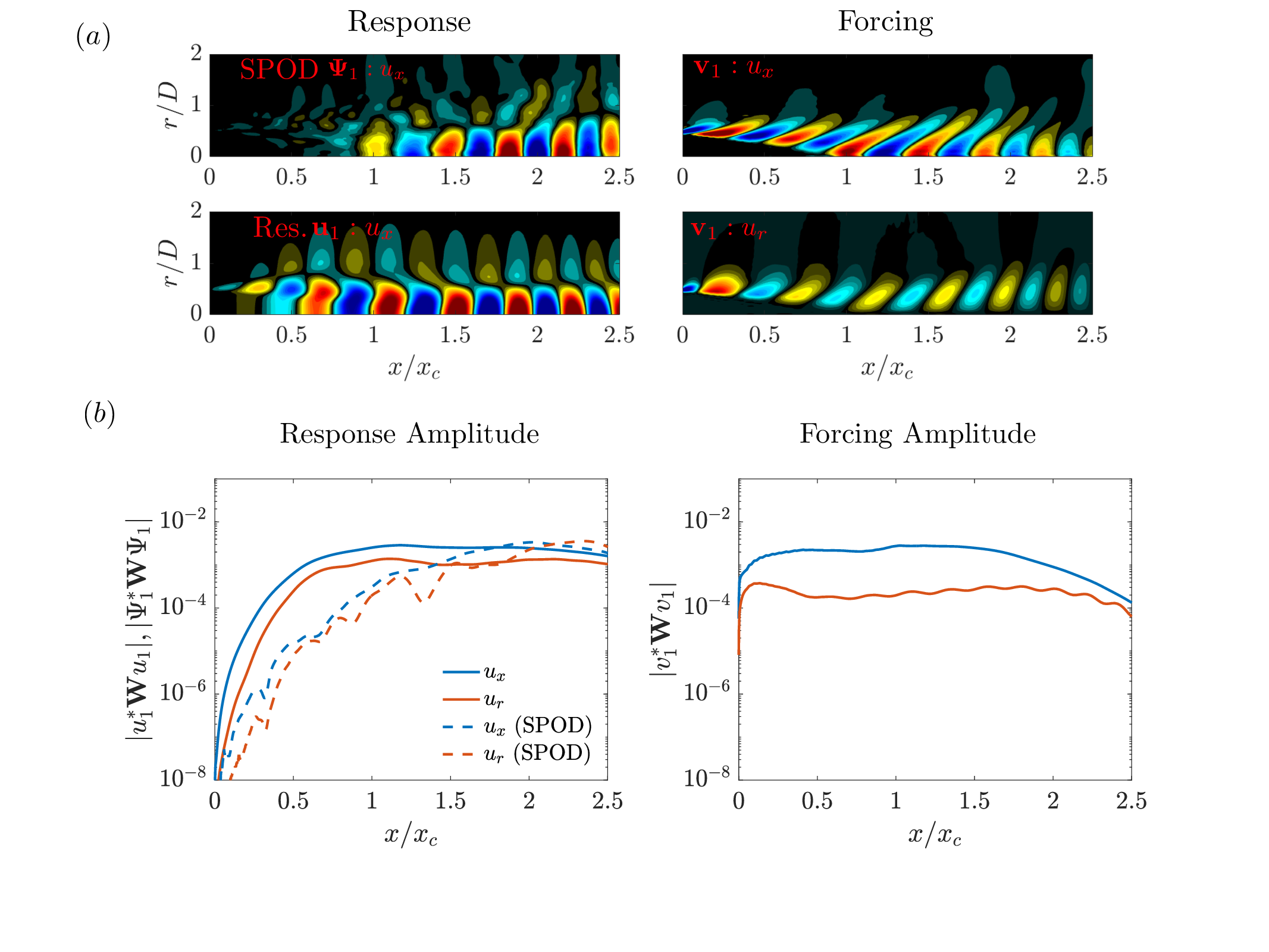}
\caption{(a) Leading forcing and response modes of the axisymmetric wavenumber $m=0$ and Strouhal number $St=0.6$ for the flight case, $M_f=0.15$. Legend is as in figure \ref{fig8}.}
\label{fig11}
\end{figure}

The Orr mechanism is dominant for the axisymmetric wavenumber and low Strouhal numbers ($St \lesssim 0.2$), where the flow dynamics are high-rank \citep{SchmidtetalJFM2018, LesshafftPFR2019}, and the KH mechanism is weak. As pointed out by \cite{PickeringJFM2020}, it also exists for $m>0$, but is overwhelmed either by streaky structures generated by the lift-up mechanism in the $St \to 0$ limit, or by KH wavepackets at higher $St$. Figures \ref{fig10} and \ref{fig11} show forcing and response modes for $(m,St)=(0,0.2)$ without and with the flight, respectively. In static condition, the response modes grow over the first 1.5 potential core lengths. This feature is consistent with the SPOD mode, but the growth process is clearly different between model and data; the rank-1 model is not sufficient for a detailed discussion of the data. The forcing modes also present and overall growth with streamwise distance, after a slight decay in the initial region. 


The flight stream changes these trends. Instead of presenting monotonic growth, the response modes saturate around $x/x_c \approx 0.5$ and propagate with constant amplitude further downstream, as shown in figure \ref{fig10}. The forcing amplitude remains essentially constant for $1.5x_c$, in contrast with the gradual increase seen in the static case. It can be seen that, despite sharing general traits, there is a significant discrepancy between leading SPOD and response modes for both jets, as indicated by the $\beta$ metric shown in figure \ref{fig6}. Notably, the SPOD modes have a much slower spatial growth than the model. Improving the agreement between model and flow data would probably require taking the suboptimals into account, since at this Strouhal number their gain is comparable to that of the optimal mode. Here the potential core scaling provides little help in explaining the modifications produced by the flight stream (see also figure \ref{beta_f_nf}), showing that they involve other effects other than mean-flow stretching.

\subsection{Lift-up mechanism}

\begin{figure}
\centering
\includegraphics[trim=0cm 0cm 3cm 0cm, clip=true,width=0.85\linewidth]{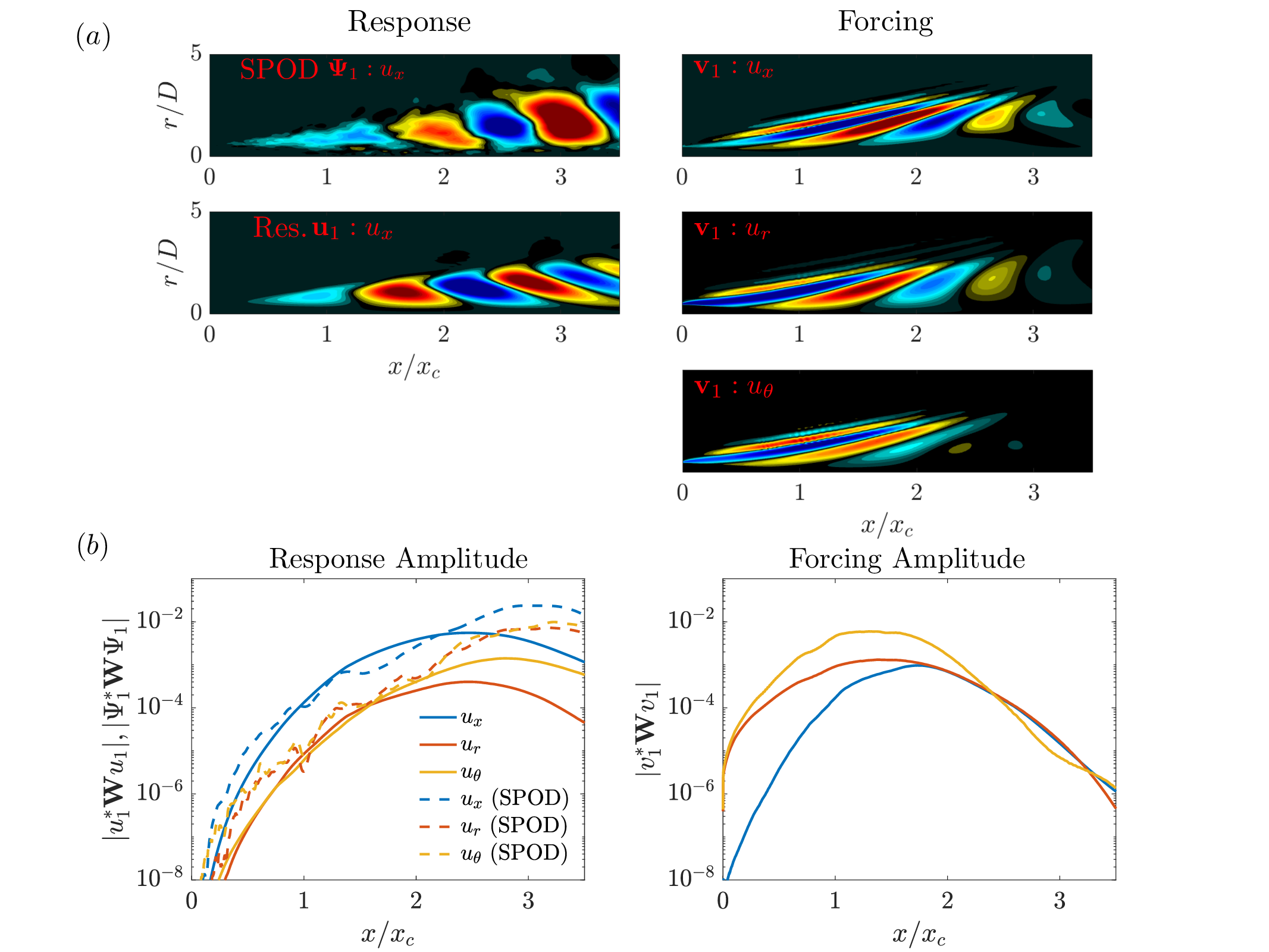}
\caption{Leading forcing and response modes of the axisymmetric wavenumber $m=3$ and Strouhal number $St=0.02$ for the static case, $M_f=0$. Legend is as in figure \ref{fig8}.}
\label{fig12}
\end{figure}

We now focus on the $St \to 0$, $m>0$ region of the frequency-wavenumber plane, whose associated coherent structures are underpinned by streaks generated via the lift-up mechanism. These structures are streamwise-elongated and forced by counter-rotating streamwise vortices. Figure \ref{fig12} shows spatial structures and amplitudes of forcing and response modes of the static case for $(m,St)=(3,0.02)$, with $St=0.02$ being the first frequency bin obtained with the FFT resolution chosen for the SPOD computation. SPOD and resolvent response modes display spatially-extended structures that follow the shear-layer development and reach their maximal amplitudes far downstream. In the static case, SPOD and resolvent modes display some similar features. For instance, their amplitude envelopes exhibit the same streamwise increase behaviour, and their wavelength is roughly the same. However, far downstream of the end of the potential core, $x/x_c>2.5$, the wavelengths of the flow structures in the SPOD and resolvent modes differ, which explains their poor alignment, as seen in figure \ref{fig6}. Forcing structures for the three velocity components are spatially-extended and inclined with respect to the mean flow, similar to an Orr-type behaviour. However, inspection of the model amplitude curves shows that the the lift-up mechanism is dominant. The radial and azimuthal components of the forcing are orders of magnitude higher than the streamwise component in the initial jet region (although this difference disappears further downstream), as opposed to the comparable contributions of the streamwise and radial components that characterise the Orr mechanism. The radial and azimuthal components form the streamwise rolls that optimally force the flow, producing positive and negative regions of fluid \enquote{lifting} via the streamwise component. As a result, the streamwise velocity component becomes dominant in the flow response, as can be observed in the response amplitude curve.

\begin{figure}
\centering
\includegraphics[trim=0cm 0cm 3cm 0cm, clip=true,width=0.85\linewidth]{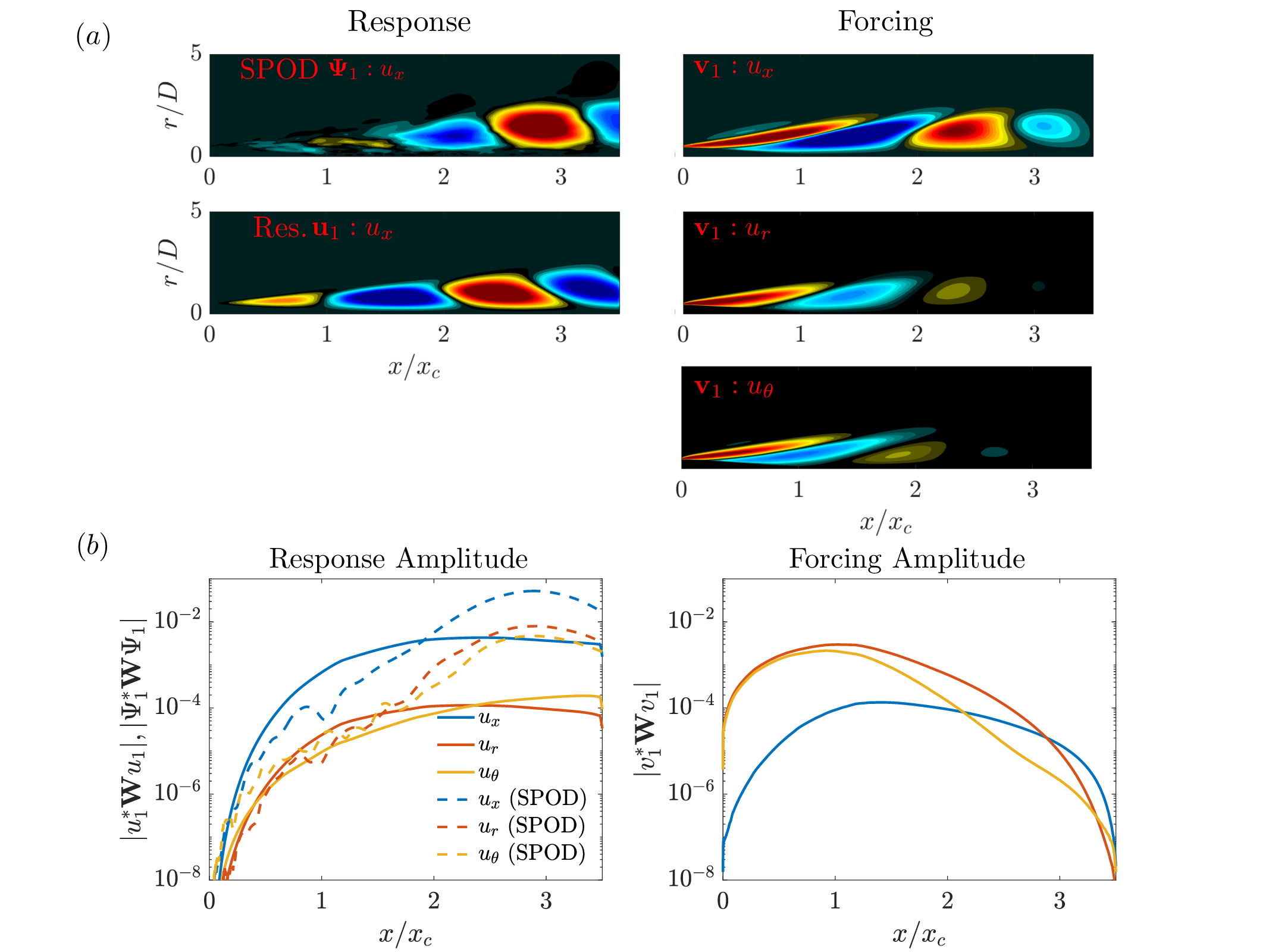}
\caption{Leading forcing and response modes of the axisymmetric wavenumber $m=3$ and Strouhal number $St=0.02$ for the flight case, $M_f=0.15$. Legend is as in figure \ref{fig8}.}
\label{fig13}
\end{figure}

Figure \ref{fig13} presents $m=3$ streaky structures in flight condition. The SPOD mode reveals a structure with larger wavelength with respect to the static case, which is a consequence of the higher convection velocity produced by the flight stream. This behaviour is correctly captured by the model, which displays larger wavelengths both in the forcing and response modes. Note that with the flight stream the SPOD and resolvent modes are clearly in better agreement, as also indicated by the $\beta$ metric shown earlier. Analysis of the forcing amplitude envelope reveals   a more marked dominance of the radial and azimuthal velocity components over the streamwise component. While that is the case in the first 1.5 potential core lengths in the static case, it occurs for approximately 2.5$x_c$ with the flight stream, suggesting clearer and stronger rolls in the forcing mechanism in flight condition. As a consequence, 
the streamwise component is reinforced in the response, and its separation to the other two components increases with respect to the static case. These trends suggest that, despite the global weakening of the linear mean-flow growth mechanisms in flight condition, due to the reduction in shear, there is a relative reinforcement of the lift-up mechanism, which becomes more clearly distinguishable in the optimal forcing and response modes. The amplitude envelopes of the SPOD modes also exhibit a larger dominance of the streamwise component, showing that the relative reinforcement of the lift-up mechanism predicted by the model is also manifest in the flow. The flight stream response modes also display slower beyond $x/x_c \approx 2$, indicating streaks that remain energetic for longer streamwise distances. Notice that, also for the streaks, the potential core scaling alone is not sufficient to correct for the modifications of the flight stream case. 

\begin{figure}
\centering
\includegraphics[trim=0cm 4cm 0cm 3cm, clip=true,width=\linewidth]{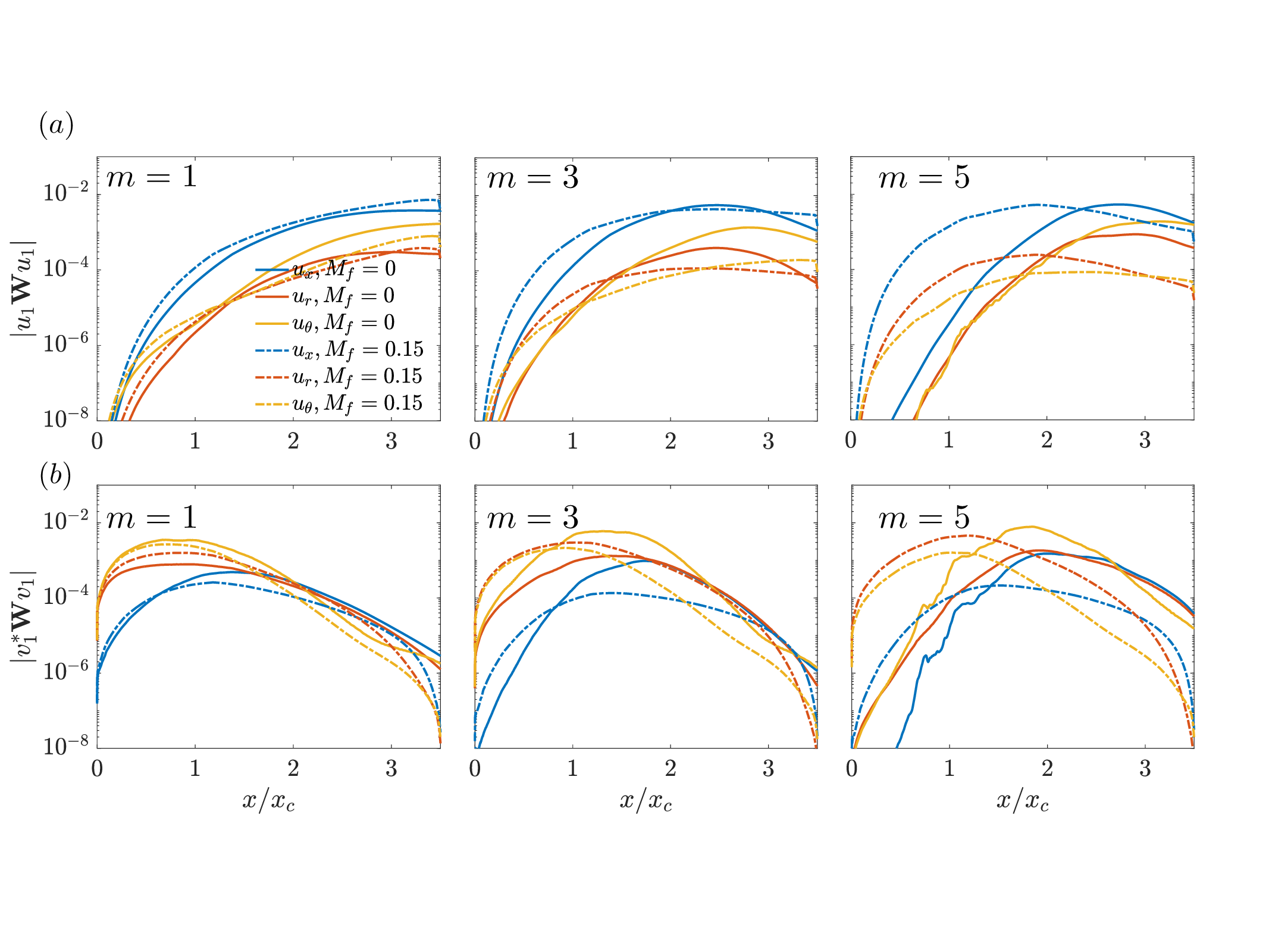}
\caption{Component-wise amplitude of leading response (a) and forcing (b) modes at $St=0.02$ for different azimuthal wavenumbers.}
\label{fig14}
\end{figure}

All of these trends were observed for other azimuthal wavenumbers in the $St \to 0$ limit. Figure \ref{fig14} shows a direct comparison between amplitudes of forcing and response modes in the static and flight cases for $m=1,3,5$. For all azimuthal wavenumbers, the flight stream leads to a more pronounced predominance of streamwise rolls in the optimal forcing. In the associated leading response modes, the separation between the streamwise component and the $u_r$-$u_\theta$ (which compose the streamwise response rolls) components is systematically larger than in static condition. It is important to emphasise that, although they are not dominant, the Orr and KH mechanism are also active at low frequencies, and the associated coherent structures for $m>0$ are likely a mixture of Orr structures, streaks and weak KH wavepackets. The results presented above suggest that in flight condition, in spite of the overall energy attenuation, the Lift-up mechanism is strengthened with respect to the other two. This is a direct consequence of the rank decrease at $St \to 0$ shown in figure \ref{fig5}. As the resolvent model predicts a larger gain separation, the more discernible streaky structures of the leading response modes are also more likely to be excited by the nonlinear forcing, and therefore more likely to be observed in the flow data. This is consistent with the improved alignment between leading SPOD and resolvent modes obtained in the flight case (figure \ref{fig6}), and with the larger eigenvalue separation observed in the SPOD (figure \ref{fig4}).

\begin{figure}
\centering
\includegraphics[trim=1cm 1cm 1cm 1cm, clip=true,width=\linewidth]{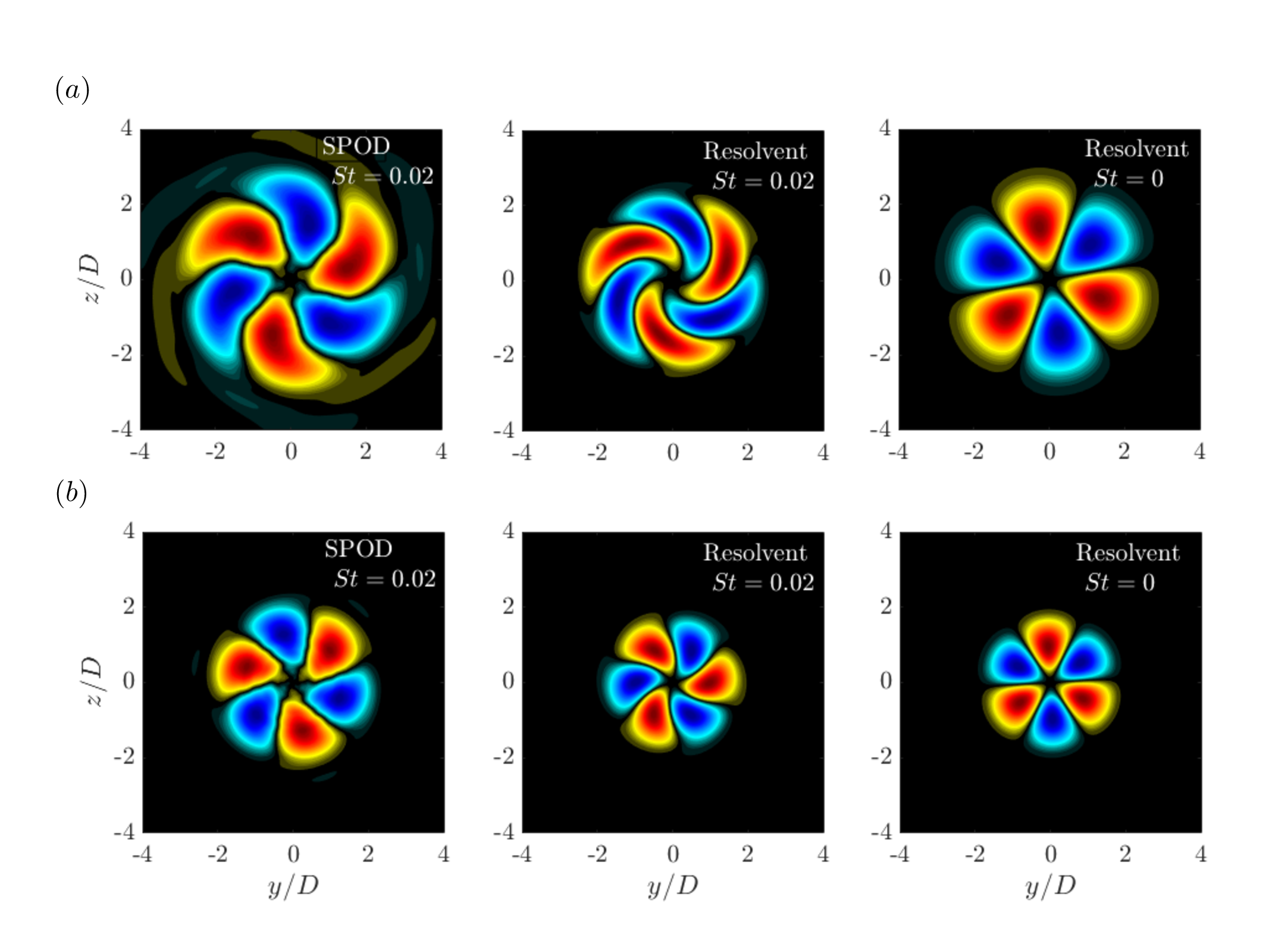}
\caption{Cross-plane cut of leading SPOD and resolvent modes at $m=3$ and taken at $x/x_c=2.5$. The real part of the streamwise velocity modes are shown, with contours scaled to $\pm||u_{x}||$. (a): static case, $M_f=0$. (b): flight case, $M_f=0.15$.}
\label{fig15}
\end{figure}

The explanation for the larger gain separation and clearer dominance of the streak mechanism comes from the larger convection velocities, and smaller associated streamwise wavenumbers, in the flight stream case. Strictly speaking, streaky structures are characterised by \textit{zero} streamwise wavenumbers, developing parallel to the jet axis. This is the case at $St=0$. For small, but non-zero frequencies, the flow structures still bear most of the chacteristics of streaks \citep{PickeringJFM2020}, but they acquire an azimuthal convection velocity, $U_\theta = \omega/m$, which makes them rotate slightly around the jet axis as they evolve downstream. This can be seen in figure \ref{fig15}, which shows a cross-plane cut, made at $x/x_c =2.5$, of $m=3$ SPOD and resolvent modes at $St=0.02$. Resolvent modes computed at $St=0$ are also shown for comparison. In the flight stream case, the streamwise convection velocity, $U_c = \omega/\alpha$, with $\alpha$ the streamwise wavenumber, at a given frequency is higher. The total convection velocity, given as the sum of the azimuthal and streamwise component, is then more aligned with the streamwise direction. Notice how, in the flight stream case, the rotation/swirling effect is much less marked. The flow structures are consequently more similar to a \enquote{classic} $St=0$ streak.

The effect of the smaller wavelength/higher convection velocity on the gain separation can be further analysed using a locally-parallel model, which takes $\alpha$ as input. In appendix \ref{appC} we use the model explored by \cite{Maia_etal_flight_2023} to show that, at $St=0$, reducing the wavelength (i.e. approaching classic streak characteristics), indeed leads to a higher non-normality/gain separation. This agrees with the results obtained in the global analyses and explains the better agreement with SPOD modes in the flight stream case.

\section{Conclusions}
\label{sec:conclusions}

We study coherent structures in turbulent subsonic jets subject to a uniform external flight stream. This work builds on the recent study of \cite{Maia_etal_flight_2023}, which presented a comprehensive characterisation of the frequency-wavenumber energy spectrum in flight condition using time-resolved PIV and high-fidelity LES databases. Here we extend their analysis by modelling coherent structures educed from the flow with global resolvent analysis. SPOD is used to characterise empirically the effect of the flight stream on the most energetic flow structures. The mode energies and spatial structures are systematically compared with gains and shapes of resolvent response modes. The alignment between SPOD and resolvent modes is high for a broand range of Strouhal number and azimuthal wavenumbers, thanks to the use of the mean-flow-consistent eddy-viscosity model in the linear operator. The model is found to correctly describe a number of important effects of flight on the jet dynamics, and it is demonstrated that the most prominent modifications are associated with linear mean-flow mechanisms, rather than the effects of non-linear interactions. Three important effects are: i) Both the frequency-azimuthal wavenumber distribution of SPOD modal energy and the change in this distribution with flight is correctly mirrored by the distribution of resolvent gains. These distributions show how low-frequency streaky/Orr stuctures, that carry the bulk of the fluctuation energy, are those most damped by the fight stream. The trend is faithfully reproduced by the resolvent model. 

ii) At intermediate Strouhal numbers, $0.4 \lesssim St \lesssim 0.8$, the low-rank behaviour of the flow is degraded by the flight stream, as evidenced by the smaller eigenvalue separation between leading and second SPOD modes. A similar reduction in gain separation is observed in the resolvent spectrum. This effect is enhanced when the computation domain is truncated to the first 1, 1.5 potential core lengths, which highlights the region where the KH mechanism is active.

iii) In the $St \to 0$ frequency limit, SPOD shows an enhanced low-rank behaviour in flight condition, despite the large attenuation of the dominant streaky structures: the low-frequency dynamics in flight are less energetic but more organised than they are in static conditions.. This behaviour is reproduced by the resolvent model. Analysis of the response modes and their component-wise amplitude curves reveals that the lift-up mechanism is more marked with the flight stream. Streamwise vortices emerge more clearly in the forcing modes with respect to the streamwise forcing component, and streamwise velocity streaks are more marked in the flow response with respect to the radial and azimuthal components. The latter trend is also manifested in the empirical structures educed from SPOD. These results reveal that the lift-up mechanism stands out more clearly for a jet in flight. We show that this effect is associated with the higher convection velocity/smaller wavenumber of the flow structures with the flight stream, which produces low-frequency streamwise velocity structures which are more aligned with the jet axis, and are thus more similar to standard zero-frequency streaks. The smaller wavelengths also result in larger gain separations predicted by the resolvent analysis, which explains the rank decrease observed at low $St$. In summary, the results show that coherent structures associated with linear instability mechanisms are globally weakened, due to a reduction in shear; but the extra convection effect of flight nonetheless makes streaks stand out more clearly than the Orr and KH structures at low $St$.


The results described here may be used to guide future sound-source models of jets in flight. The acoustic field of such jets present broadbad changes with respect to the static case \citep{VishwanathanFlight}, which are likely associated to changes in coherent structures (and associated instability mechanisms) in the turbulent field. Resolvent analysis is shown here to be clearly equipped to reproduce, to a great extent, these changes; therefore it might provide a useful framework to explore sound-radiation mechanisms.

\section*{Acknowledgements}

This work has received funding from the Clean Sky 2 Joint Undertaking (JU) under the European Union's Horizon 2020 research and innovation programme under grant agreement No 785303. Results reflect only the authors' view and the JU is not responsible for any use that may be made of the information it contains. The LES studies were performed at Cascade Technologies and were supported in part by NAVAIR SBIR project with computational resources provided by DoD HPCMP. The authors would like to thank André Cavalieri and Diego Blanco for helpful discussions regarding streaky structures and convergence of SPOD modes at low frequency.  I.A.M. also acknowledges support from the Science Without Borders program through the CNPq Grant No. 200676/2015-6.

\appendix

\section{Trapped waves in the potential core}\label{appA}

\begin{figure}
\centering
\includegraphics[trim=1cm 10.5cm 0.5cm 8cm, clip=true,width=\linewidth]{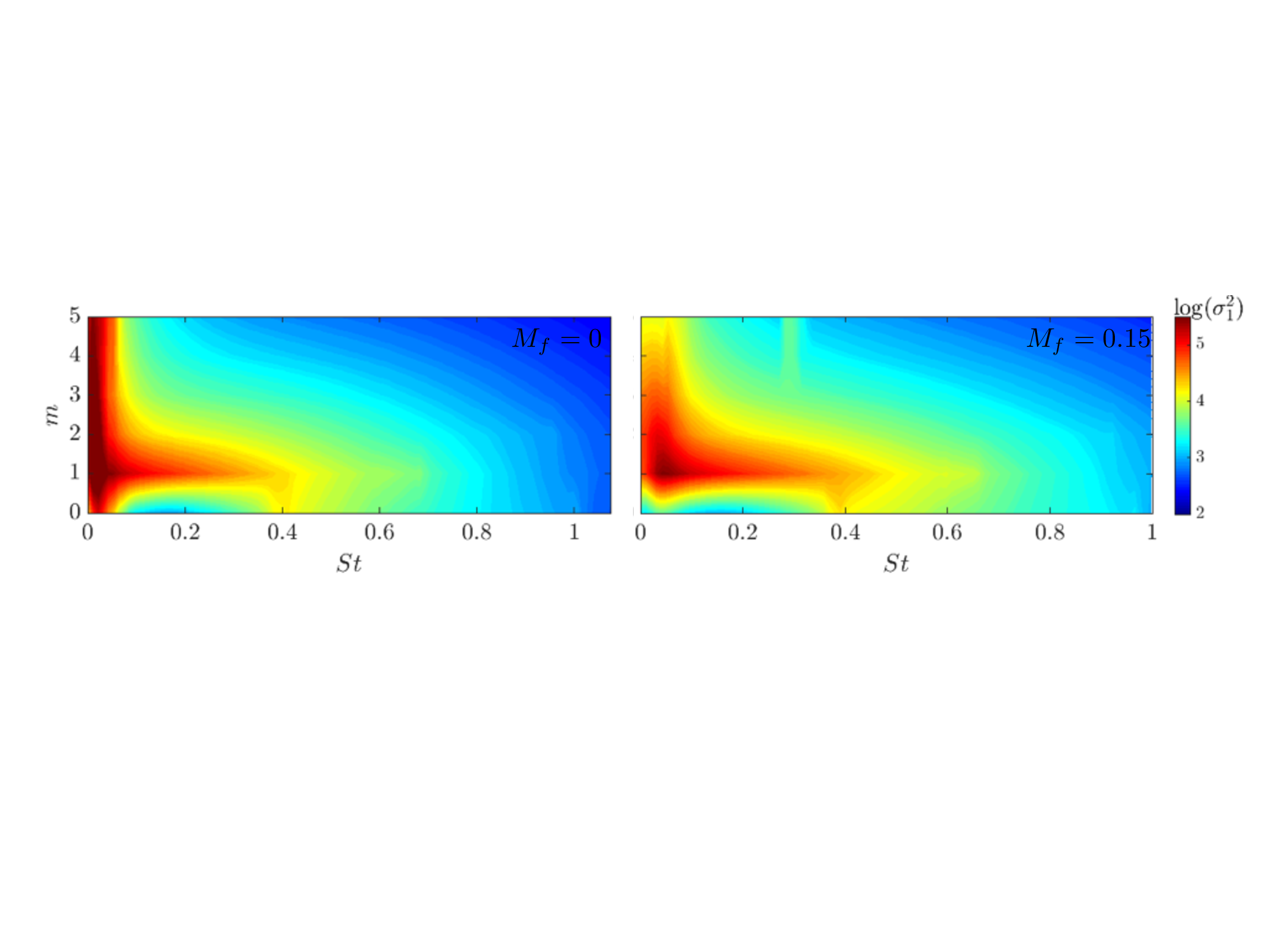}
\caption{Leading resolvent gain, $\sigma_1^2$, computed without the potential core mask. Contours are in $\mathrm{log}_{10}$ scale.}
\label{fig16}
\end{figure}

This appendix presents results of resolvent analyses carried out with the mean-flow mask shown in figure \ref{fig1}(b) applied to both the forcing and response fields, thus removing the restriction in the forcing field inside the jet potential core. Figure \ref{fig16} shows the leading gain maps for the two flow conditions studied. Overall, the maps are quite similar to those shown in section \S \ref{sec:en_maps}, obtained with the core mask. The highest optimal amplification occurs at the $St \to 0$ limit and concerns streaky/Orr structures that dominate the flow in the jet far-field. These are the structures most affected by the flight stream, which significantly reduces their associated gains. Notice, however, the spike that emerges for the axisymmetric mode around $St=0.4$ (and that cannot be clearly seen in the SPOD modal energy maps). This frequency matches the location of branches of trapped acoustic modes in jet core \citep{schmidt_etal_2017} identified through global stability analysis.

\begin{figure}
\centering
\includegraphics[trim=3cm 7cm 4cm 4cm, clip=true,width=\linewidth]{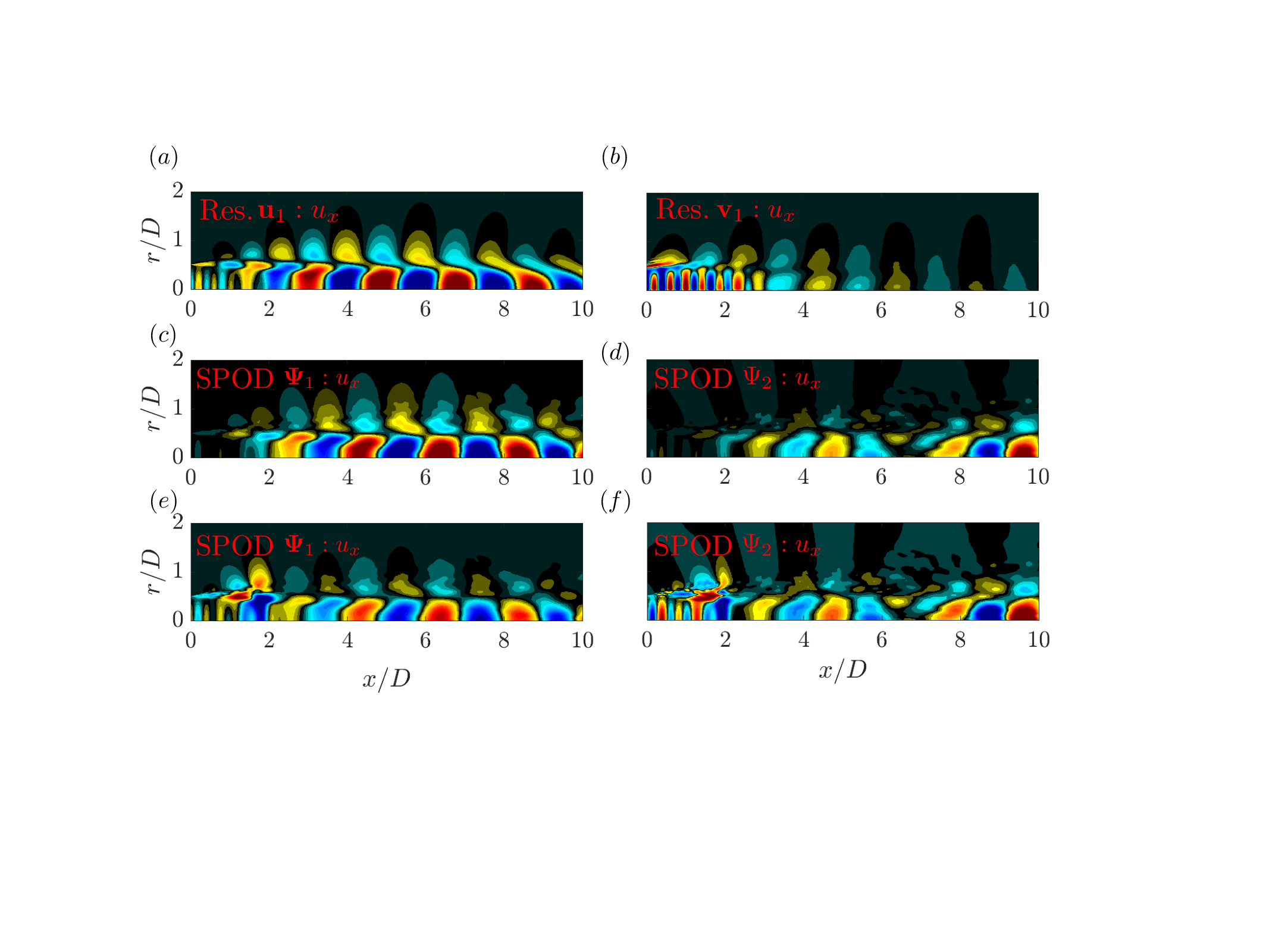}
\caption{Resolvent and SPOD modes at $St=0.4$. (a) and (b) show the leading resolvent response and forcing modes. (c) and (d) display the leading and second SPOD modes, respectively. In (e) and (f) the amplitude of the KH wavepackets are artifically reduced by 95\%, in order to highlight the trapped waves in the jet core. The streamwise velocity component is shown, with contours corresponding to $\pm 0.7||u_x||$ in (a)-(d).}
\label{fig17}
\end{figure}

The signature of these trapped waves can be clearly seen in the forcing and response modes at that Strouhal number, which are reported in figure \ref{fig17} for the $M_f=0$ jet. These waves are also found to be present in the SPOD modes; but their energy is small, and they are overwhelmed by KH wavepackets. In figures \ref{fig17}(e)-(f), the amplitude of KH wavepackets in the SPOD modes is artifically decreased by 95\% in order to highlight the trapped waves. They are clearly present in the jet core, and are more pronounced in the first suboptimal mode. Their signature, however, is not apparent in the eigenvalue spectrum.

\section{SPOD and resolvent analysis with truncated domains}\label{appB}

\begin{figure}
\centering
\includegraphics[trim=0cm 4cm 0cm 4cm, clip=true,width=\linewidth]{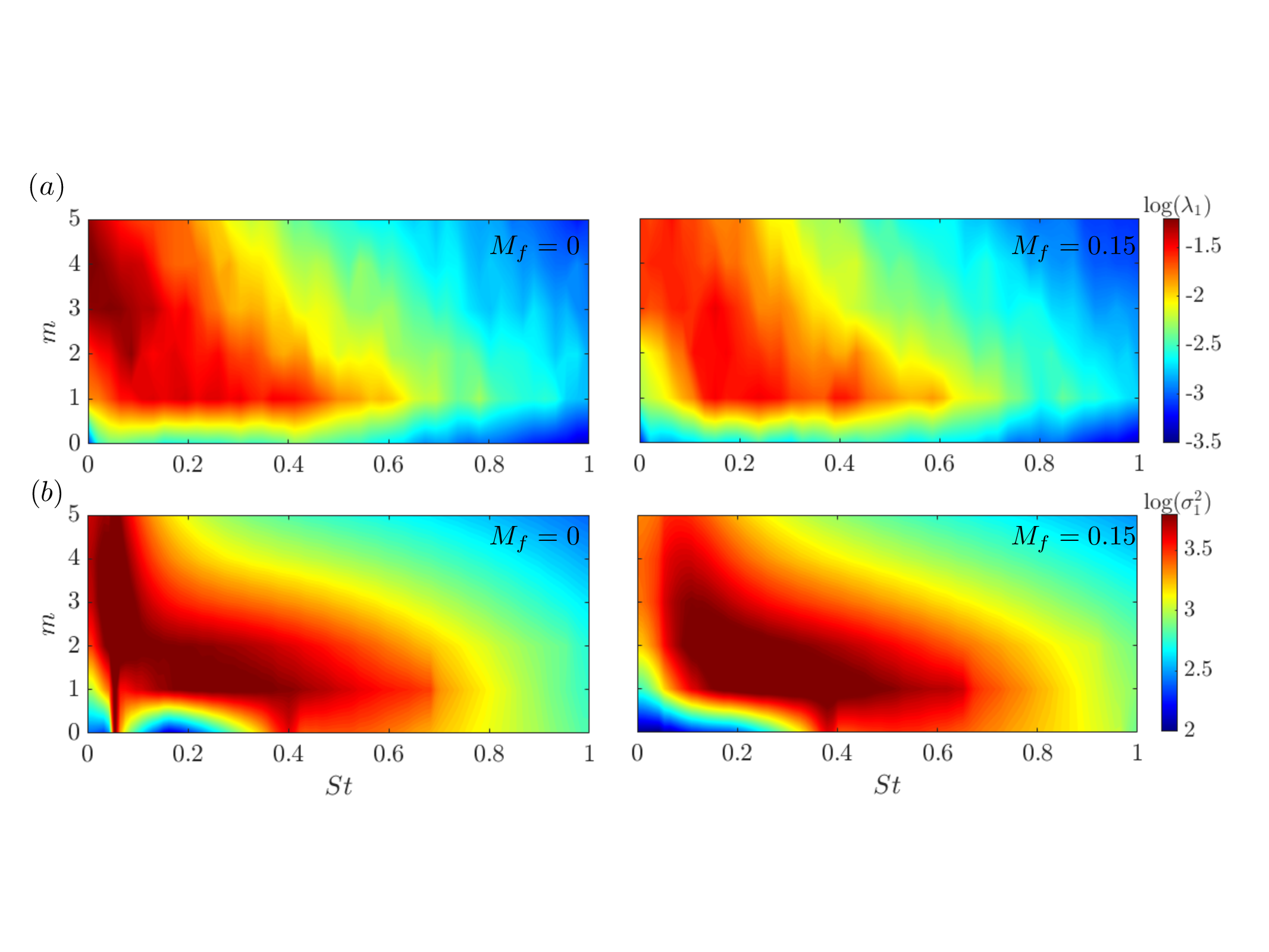}
\caption{Eigenvalue and resolvent gain spectra computed with domains truncated at $x/x_c=1$.}
\label{fig18}
\end{figure}

\begin{figure}
\centering
\includegraphics[trim=0cm 4cm 0cm 4cm, clip=true,width=\linewidth]{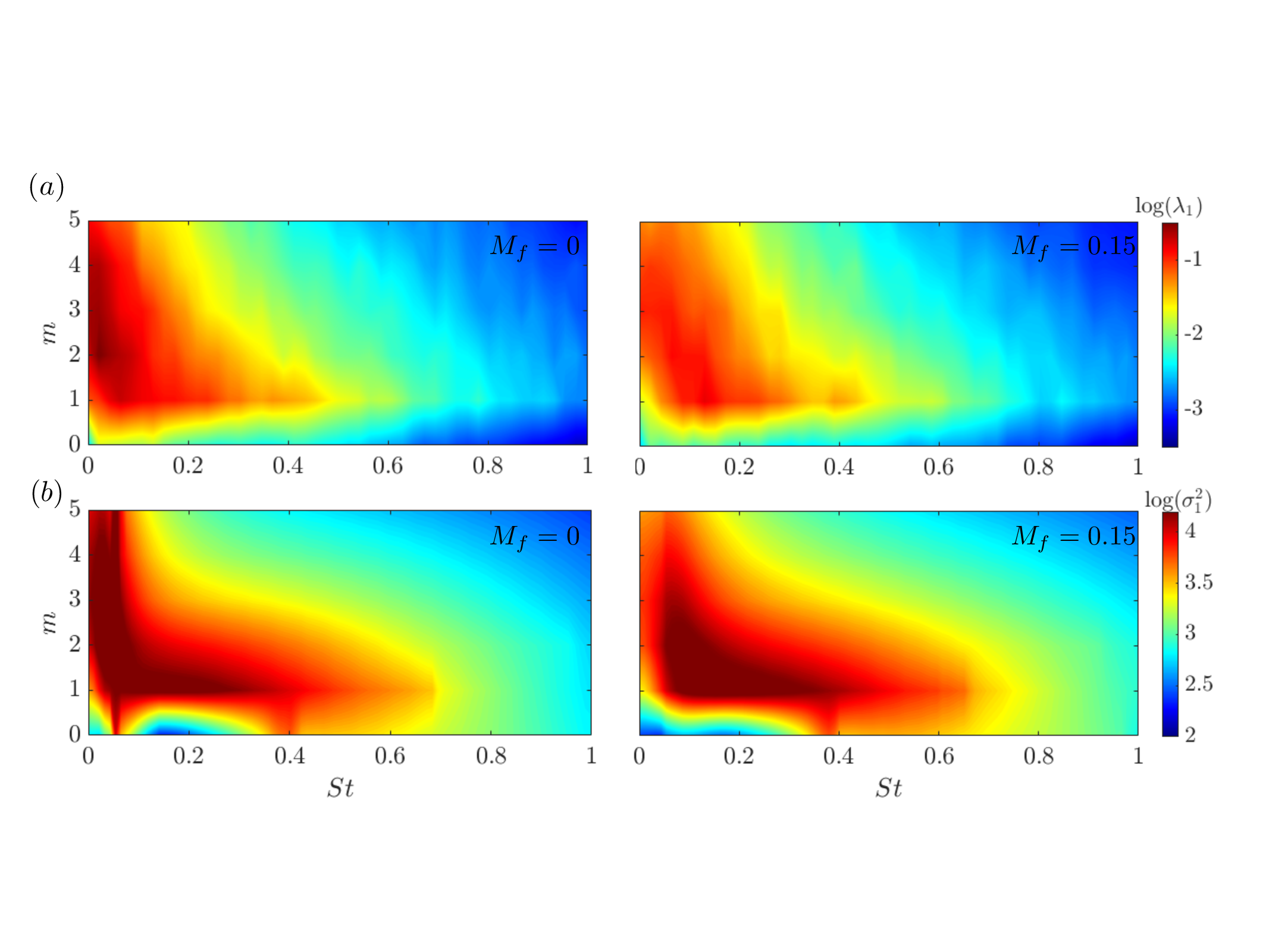}
\caption{Eigenvalue and resolvent gain spectra computed with domains truncated at $x/x_c=1.5$}
\label{fig19}
\end{figure}

When performing SPOD and resolvent analysis with the full computational domain, that extends up to 30D in the streamwise direction, modal energy and resolvent gains are biased towards low-frequency structures that dominate the jet far downstream, and that are underpinned by the non-modal Orr and Lift-up mechanisms. That inevitably masks most of the contribution of KH mechanism to the global energy/gain spectrum, as they are convectively unstable in the initial jet region, approximately up to the end of the potential core. As KH wavepackets are highly efficient acoustic radiators \citep{JordanColoniusReview}, the initial jet region is of fundamental important for understanding jet noise. In this section we present results of resolvent analyses performed with domains truncated at $x/x_c=1$ and $x/x_c=1.5$, with a view to highlighting the changes produced by the flight stream on the zones of the spectrum underpinned by modal instability. Modal energy and gain maps are shown in figures \ref{fig18} and \ref{fig19}. The spectra are much more broadband with respect to those obtained with the full domain and include, in addition to the high energy/amplification zone near the $St \to 0$ limit, considerable energy/amplification in intermediate frequencies, $0.2 \lesssim St \lesssim 0.8$. The signature of the $m=0$ wavepackets, peaking toward $St=0.4$-$0.8$ are also more clear in the truncated maps (more so in the resolvent results). The clearest effect of the flight stream is still the attenuation of the $St \to 0$ zone, as seen previously for the full domain, but with the attenuation now concentrated at higher $m$, as going upstream the peak energy evolves towards higher $m$ \citep{Maia_etal_flight_2023}.

\begin{figure}
\centering
\includegraphics[trim=3cm 0.5cm 3cm 4cm, clip=true,width=0.85\linewidth]{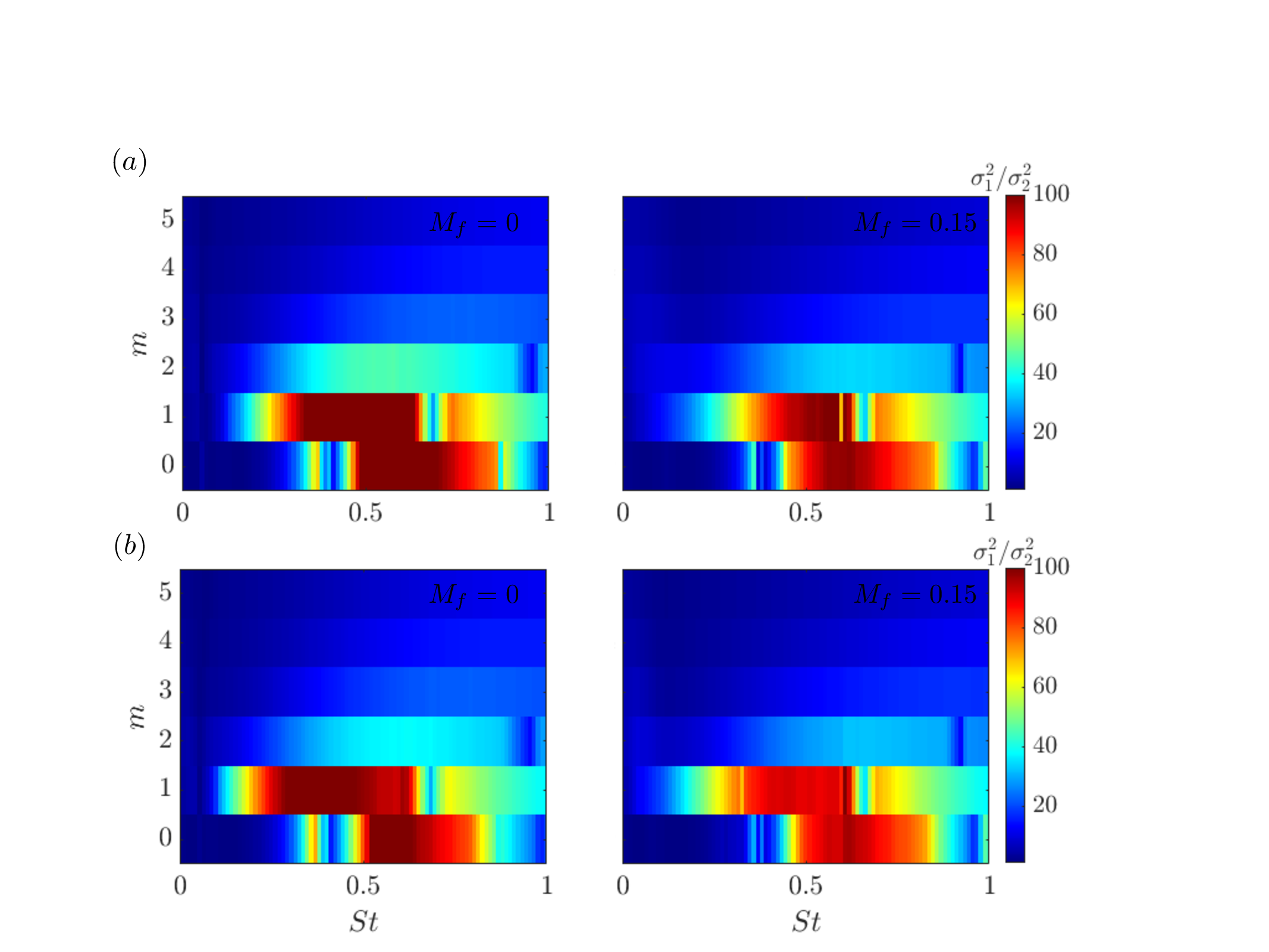}
\caption{Resolvent gain separation, $\sigma_1^2/\sigma_2^2$ computed with a domain truncated at (a): $x/x_c=1$; and (b): $x/x_c=1.5$}
\label{fig20}
\end{figure}

Figure \ref{fig20} shows the gain separation, $\sigma_1^2/\sigma_2^2$ in the truncated domain. A reduction in gain separation in flight condition becomes apparent, highlighting the weakening of the low-rank behaviour in the KH-dominated zone. This trend is also consistent with a smaller growth rate of the KH instability, whose mechanism is mainly manifest in the leading resolvent mode.

\section{Locally-parallel model: gain separation at $St=0$ and low $\alpha$}\label{appC}

In this section, we consider a locally-parallel model to explore the effect of decreasing wavenumber on the gain separation of the resolvent operator. In the locally-parallel framework, we assume flow perturbations of the form,

\begin{equation}
\mathbf{q}'(x,r,\theta,t)=\hat{\mathbf{q}}(r)\mathrm{exp}^{i(\alpha x - \omega t +m\theta)},
\label{ansatz}
\end{equation}
where the radial structure of the perturbations is given by $\hat{\mathbf{q}}(r)$, $\alpha$ and $m$ are streamwise and azimuthal wavenumbers, respectively, and $\omega$ is the frequency. Fourier-transforming the Navier-Stokes equations and introducing the above Ansatz yields,

\begin{equation}
\hat{\mathbf{q}}_{\alpha,\omega,m} = \mathcal{C}\left(i\omega\mathcal{I} -\mathcal{A}_{0} - \alpha\mathcal{A}_{1} - \alpha^2 \mathcal{A}_{2} \right)_{\bar{\mathbf{q}}}^{-1}\mathcal{B}\hat{\mathbf{f}}_{\alpha,\omega,m},
\label{ns_fft}
\end{equation}
where the linear operators $\mathcal{A}_{0}, \mathcal{A}_{1}$, and $\mathcal{A}_{2}$ contain terms issuing from zero-th, first and second order derivatives in $x$, respectively. Matrices $\mathcal{B}$ and $\mathcal{C}$ can be used to restrict forcing and response to a desired subspace. In a more compact form, we can write,

\begin{equation}
\hat{\mathbf{q}}_{\alpha,\omega,m}=\mathcal{R}_{\bar{\mathbf{q}},\alpha,\omega,m}\hat{\mathbf{f}}_{\alpha,\omega,m},
\end{equation}
where $\mathcal{R}_{\bar{\mathbf{q}},\alpha,\omega,m} = \mathcal{C}\left(i\omega -\mathcal{A}_{0} - \alpha\mathcal{A}_{1} - \alpha^2 \mathcal{A}_{2} \right)_{\bar{\mathbf{q}}}^{-1}\mathcal{B}$ is the resolvent operator. The discretisation in the radial direction is carried out using Chebyshev collocation points. The domain is extended to the far-field by mapping the original domain,  $r \in [-1, 1]$ to $r \in [0,\infty)$ using a mapping. The reader is referred to \citep{Maia_etal_flight_2023} for details about the matrices and boundary conditions. The locally-parallel resolvent analysis is carried out at a fixed streamwise position using the mean flow, $\alpha$, $\omega$ and $m$ as inputs. The mean flow profile were based on the LES data, fitted with the hyperbolic tangent profile proposed by \cite{MuchalkeHermann1982} for the static case. Here we consider a profile taken at $x/x_c=0.7$. and a Reynolds number of $Re=50$, which is consistent with the eddy viscosity model used in the global resolvent framework. Since we are interested in the behaviour of streaky structures, we set $\omega=0$ andt The azimuthal wavenumber is $m=3$.

\begin{figure}
\centering
\includegraphics[trim=0cm 0cm 0cm 0cm, clip=true,width=0.5\linewidth]{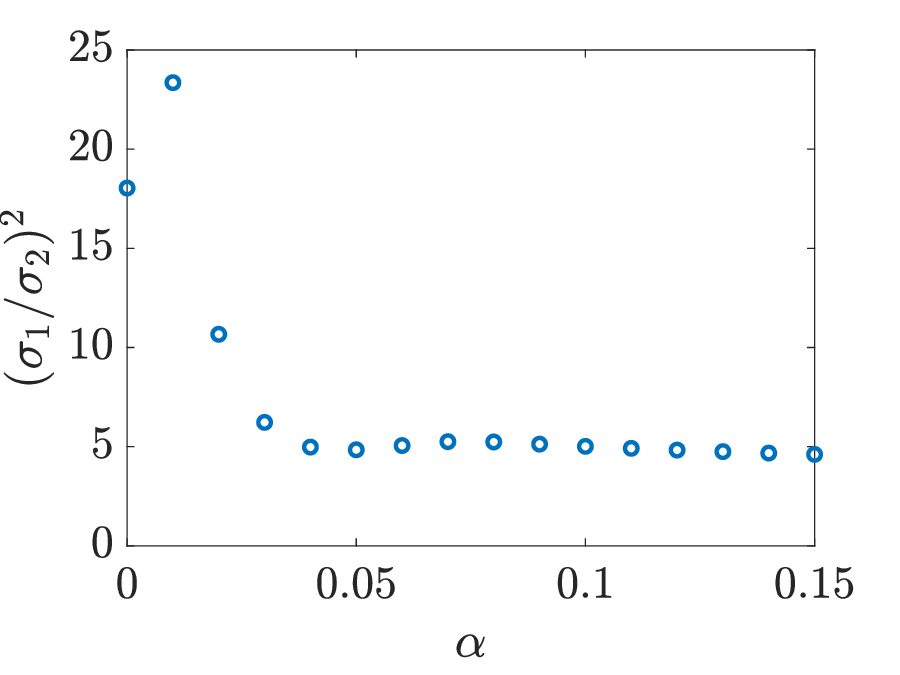}
\caption{Resolvent gain separation, $(\sigma_1/\sigma_2)^2$ computed with a locally-parallel model for $St=0$ and low values of the streamwise vawenumber, $\alpha$.}
\label{fig21}
\end{figure}

Figure \ref{fig21} shows the resolvent gain separation computed for different values of $\alpha$. It can be observed that, although the larger gain separation is not strictly at $\omega=0$, approaching zero streamwise wavenumber (\enquote{classic} streaky structures), can lead to substantially larger gain separations. This explains why this behaviour observed in the global framework, and why it leads to a clearly distinguishable streak mechanism in the forcing and response modes.

\bibliographystyle{plainnat}
\bibliography{bibfile}

\end{document}